\documentclass[journal,twoside,web]{ieeecolor}
\usepackage{tmi}
\usepackage{cite}

\usepackage{amsmath,amssymb,amsfonts}
\usepackage{algorithmic}
\usepackage{graphicx}
\usepackage{textcomp}
\usepackage{color}
\usepackage{caption}
\usepackage{subcaption}

\usepackage{multirow} 
\usepackage{xcolor}
\usepackage{hyperref}

\definecolor{myCOLOR}{rgb}{0,0,0}
\definecolor{myCOLOR2}{rgb}{0,0,0}

\def\BibTeX{{\rm B\kern-.05em{\sc i\kern-.025em b}\kern-.08em
		T\kern-.1667em\lower.7ex\hbox{E}\kern-.125emX}}
\begin{document}
	\title{Spach Transformer: Spatial and Channel-wise Transformer Based on Local and Global Self-attentions for PET Image Denoising}
	\author{Se-In Jang, Tinsu Pan,  Ye Li,  Pedram Heidari,  Junyu Chen,  Quanzheng Li, and Kuang Gong
		\thanks{This work was supported by the National Institutes of Health under grants R03EB030280, R21AG067422, R01AG078250 and P41EB022544. Corresponding author: Kuang Gong (email: KGONG@bme.ufl.edu).}
		\thanks{S.~Jang,  Q.~Li and K.~Gong are with Gordon Center for Medical Imaging and Center for Advanced Medical Computing and Analysis , Massachusetts General Hospital and Harvard Medical School,  Boston, MA, USA }
		\thanks{K.~Gong is also with J. Crayton Pruitt Family Department of Biomedical Engineering, University of Florida, Gainesville, FL, USA}		
		\thanks{T. ~Pan is with Department of Imaging Physics, University of Texas MD Anderson Cancer Center, Houston, TX, USA}
		\thanks{Y.~Li is with Center for Advanced Medical Computing and Analysis , Massachusetts General Hospital and Harvard Medical School,  Boston, MA, USA}
		\thanks{P.~Heidari is with Division of Nuclear Medicine and Molecular Imaging,  Massachusetts General Hospital and Harvard Medical School,  Boston, MA, USA}
		\thanks{J.~Chen is with Department of Electrical and Computer Engineering and Department of Radiology, Johns Hopkins University,  Baltimore,  MD,  USA 
		}
	}

	\maketitle
	
	\begin{abstract}
		
		Position emission tomography (PET) is widely used in clinics and research due to its quantitative merits and high sensitivity, but suffers from low signal-to-noise ratio (SNR).  Recently convolutional neural networks (CNNs) have been widely used to improve PET image quality.  Though successful and efficient in local feature extraction, CNN cannot capture long-range dependencies well due to its limited receptive field.  Global multi-head self-attention (MSA) is a popular approach to capture long-range information. However, the calculation of global MSA for 3D images has high computational costs. In this work, we proposed an efficient spatial and channel-wise encoder-decoder transformer,  Spach Transformer, that can leverage spatial and channel information based on local and global MSAs.  Experiments based on datasets of different PET tracers,  i.e.,  $^{18}$F-FDG,  $^{18}$F-ACBC,  $^{18}$F-DCFPyL, and $^{68}$Ga-DOTATATE, were conducted to evaluate the proposed framework.  Quantitative results show that the proposed Spach Transformer framework outperforms state-of-the-art deep learning architectures. Our codes are available at \href{https://github.com/sijang/SpachTransformer}{https://github.com/sijang/SpachTransformer}
		
	\end{abstract}
	
	
	\section{Introduction}
	\label{sec.introduction}
	\IEEEPARstart{P}{ositron} emission tomography (PET) is an imaging modality widely used in oncology, neurology, and cardiology studies  \cite{sweet1951uses, cherry2018total, fletcher2008recommendations, beyer2000combined, di2007clinical, el2011improvement}.  It can observe molecular-level activities {\it{in vivo}} through the injection of specifically designed radioactive tracers.  Due to various physical degradation factors and limited counts received,  the signal-to-noise ratio (SNR) of PET is inferior to other imaging modalities, which compromises its clinical values in diagnosis, prognosis, staging and treatment monitoring.  Additionally,  to improve the hospital's throughput or reduce the radiation exposures to patients,  faster PET imaging or low-dose PET imaging is desirable,  where the counts received during the scan is even less.  This further challenges our ability to attain high-quality PET images from limited counts.
	
	For the past decades, various post-processing methods have been investigated to further improve PET image quality. Gaussian filtering is widely used in clinical scanners to reduce the noise but it can also smooth out image structures. Non-local mean  \cite{dutta2013non}, wavelet \cite{boussion2009incorporation}, highly constrained back-projection processing \cite{christian2010dynamic}, and guided filtering \cite{yan2015mri} have been proposed to further preserve image details based on pre-defined filters or similarity calculation.  Recently,  the convolutional neural network (CNN)-based approaches have become a dominant trend in various PET research topics such as image reconstruction \cite{kim2018penalized,gong2018iterative,gong2018petre,gong2021direct} and denoising \cite{ gong2018pet, cui2019pet, zhou2020supervised, geng2021content},  due to their better performance than other state-of-the-art methods.  The CNN-based Unet architecture \cite{ronneberger2015u} is the most widely used network architecture currently.  Performance of CNN for PET image denoising can be further improved by different loss-function designs,  e.g.,  a perceptual loss \cite{gong2018pet} formed by a pre-trained network or a learned discriminative loss from the additional discriminator network \cite{goodfellow2014generative,zhou2020supervised}.
	
	One pitfall of CNN is that it specifically focuses on local spatial information and thus has a limited receptive field.  The multi-head self-attention (MSA)-based structure,  i.e., transformer,  has the ability to capture long-range information \cite{vaswani2017attention, dosovitskiy2020image, han2022survey}.  Based on the aggregation of tokens, global MSAs were first developed for natural language processing \cite{vaswani2017attention}.  Vision Transformer (ViT) was then proposed where global MSAs were successfully applied to sequences of spatial image patches for image classification  \cite{dosovitskiy2020image}.  Due to global MSA's quadratically growing computational complexity along the spatial dimension,  Swin Transformer \cite{liu2021swin} was proposed to efficiently calculate local MSAs using shifted windows that offered linear computational complexity.  Swin Transformer calculated local MSAs on multiple stages and achieved better classification performance than global MSAs. However, local MSAs of Swin Transformer is still limited regarding the receptive field of MSA. Restormer \cite{zamir2021restormer} was recently proposed to efficiently compute global MSAs along the channel dimension with linear computational complexity for image restoration tasks.  Regarding medical imaging, various transformer-based architectures were investigated for medical image segmentation \cite{chen2021transunet,  xie2021cotr, wang2021transbts, zhang2021transfuse, hatamizadeh2022unetr, chen2021transmorph}.  For PET image denoising, transformer-based architectures have not yet been fully investigated.
	
	\begin{figure}[!t]
		\centerline{\includegraphics[width=\columnwidth]{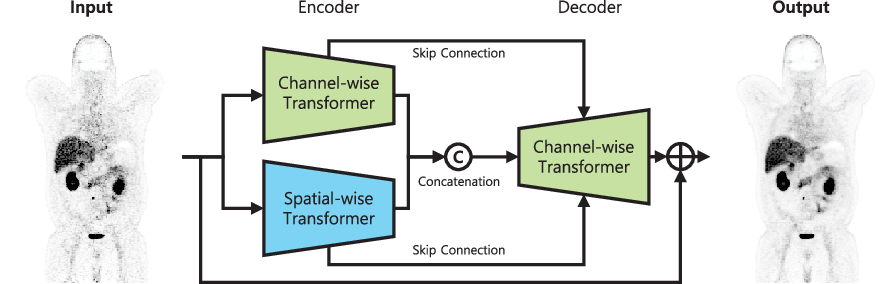}}
		\caption{A brief overview of Spach Transformer for PET image denoising.}
		\label{fig.proposed}
	\end{figure}
	\begin{figure*}[!t]
		\centerline{\includegraphics[width=17cm]{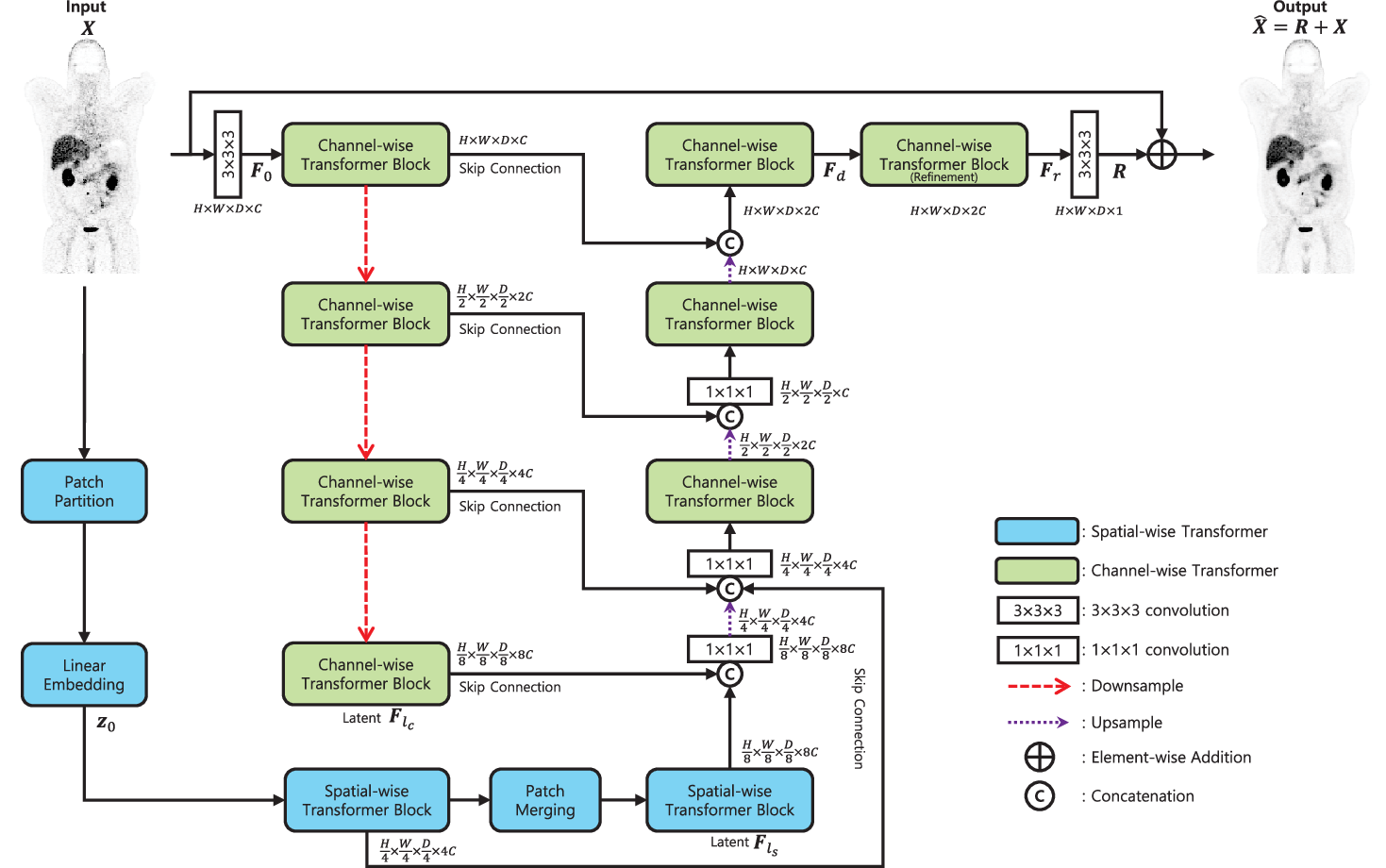}}
		\caption{Diagram of the proposed Spach Transformer for PET image denoising.  Details of the channel-wise transformer block and the spatial-wise transformer block are given in Fig.~\ref{fig.sw} and Fig.~\ref{fig.cw}, respectively.}
		\label{fig.proposed2}
	\end{figure*}
	
	In this work, we proposed a {\it{spa}}tial and {\it{ch}}annel-wise encoder-decoder {\it{transformer}},  denoted as {\it{Spach Transformer}}, that can leverage spatial and channel information from oncological PET images.  The proposed Spach Transformer includes spatial-wise and channel-wise transformer blocks,  which can efficiently perform image denoising while also preserving image details, e.g., small lesion uptakes.  The spatial-wise block is based on local attention and the channel-wise block is based on global attention.  As the channel-wise transformer block may ignore the spatial information, a gated-conv feed-forward network (GCFN) block was further designed to incorporate spatial attributes between layers in the channel-wise transformer block.  A brief overview of the proposed Spach Transformer is presented in Fig. \ref{fig.proposed}.  It encoded the input and built two deep latent features in spatial and channel directions.  The decoder further took the two latent features concatenated as input to generate the final denoised PET image based on skip connections. To evaluate the effectiveness of the proposed Spach Transformer for PET image denoising,  clinical $^{18}$F-FDG,  $^{18}$F-ACBC,  $^{18}$F-DCFPyL, and $^{68}$Ga-DOTATATE datasets were utilized to quantitatively compare the performance of different CNN and transformer-based methods. 
	
	The main contributions of this paper include : (1) proposing a novel spatial and channel-wise encoder-decoder transformer integrating local and global attentions for PET image denoising and tumor-uptake preserving; (2)  designing a GCFN block to incorporate spatial attributes in the channel-wise transformer block to boost the performance; (3) evaluations using clinical datasets based on different PET tracers. This paper is organized as follows.  Section 2 presents the related works and the proposed Spach Transformer in detail.  Experiments, results and discussions are shown in Sections 3, 4, and 5, respectively.  Finally,  conclusions are drawn in Section 6.
	
	\begin{figure}[t]
		\centering
		\begin{subfigure}[b]{\columnwidth}
			\centering
			\includegraphics[width=\columnwidth]{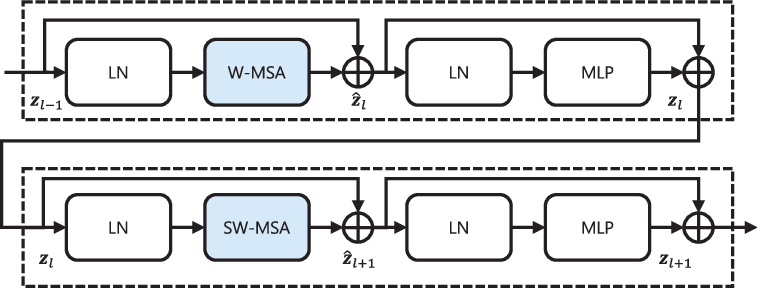}
			\caption{The spatial-wise transformer block used in this work.}
			\label{fig.sw_block}
		\end{subfigure}
		\hfill
		\begin{subfigure}[b]{\columnwidth}
			\centering
			\includegraphics[width=\columnwidth]{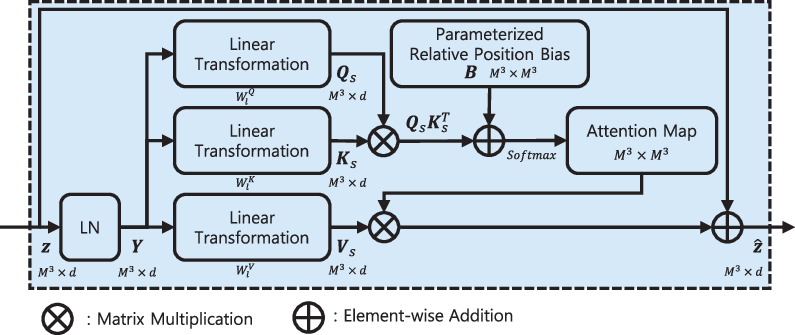}
			\caption{The window-based multi-head self-attention.}
			\label{fig.sw_sa}
		\end{subfigure}
		\caption{Diagrams of the spatial-wise transformer block (a) and the window-based attention calculation (b). }
		\label{fig.sw}
	\end{figure}
	\begin{figure}[t]
		\centering
		\begin{subfigure}[b]{\columnwidth}
			\centerline{\includegraphics[width=5.5cm]{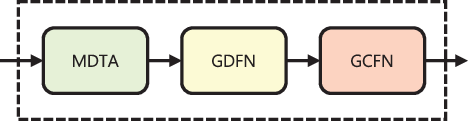}}
			\caption{The proposed channel-wise transformer block.}
			\label{fig.cw_block}
		\end{subfigure}
		\hfill
		\begin{subfigure}[b]{\columnwidth}
			\centerline{\includegraphics[width=\columnwidth]{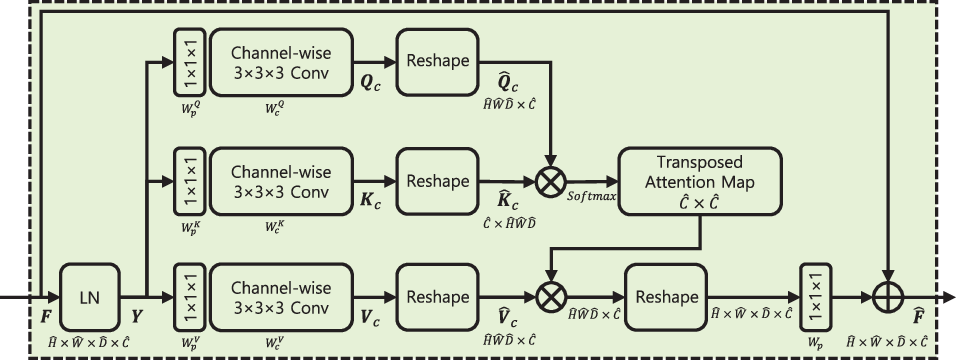}}
			\caption{The multi-dconv head transposed attention (MDTA).}
			\label{fig.cw_mdta}
		\end{subfigure}
		\hfill
		\begin{subfigure}[b]{\columnwidth}
			\centerline{\includegraphics[width=7cm]{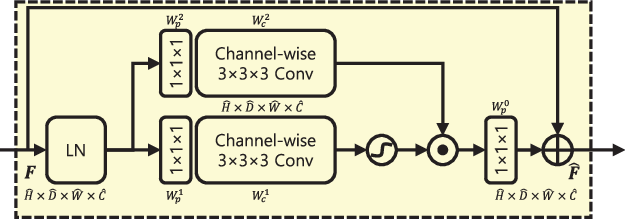}}
			\caption{The gated-dconv feed-forward network (GDFN).}
			\label{fig.cw_gdfn}
		\end{subfigure}
		\hfill
		\begin{subfigure}[b]{\columnwidth}
			\centerline{\includegraphics[width=7cm]{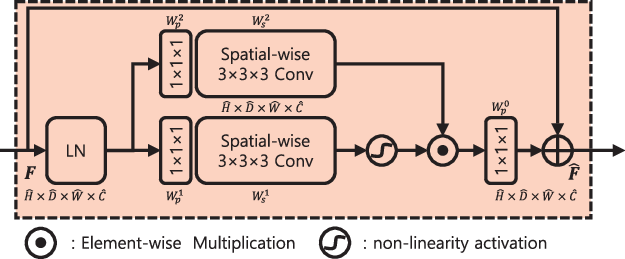}}
			\caption{The gated-conv feed-forward network (GCFN).}
			\label{fig.cw_gcfn}
		\end{subfigure}
		\caption{Diagrams of the proposed channel-wise transformer block (a) and the MDTA (b), GDFN (c), and GCFN (d) modules.}
		\label{fig.cw}
	\end{figure}
	
	\section{Method} \label{sec.proposed}
	Fig. \ref{fig.proposed2} shows the detailed diagram of the proposed Spach Transformer.  The encoder part of Spach Transformer consists of {\color{black}two levels of spatial-wise transformer blocks and four levels of channel-wise transformer blocks} {\textcolor{myCOLOR}{to fit each other's dimensionality.}}  The decoder part of Spach Transformer includes {\color{black}four levels of channel-wise transformer blocks}.  Below we will describe major elements of the Spach Transformer (Sec.~\ref{sec.proposed}) as well as details of the spatial-wise transformer block (Sec.~\ref{sec.spatial}) and the channel-wise transformer block. (Sec.~\ref{sec.channel}).  
	
	\subsection{Spach Transformer: Spatial and Channel-wise Transformer} \label{sec.proposed}
	Given a low-dose PET image ${\bf X} \in \mathbb{R}^{H\times W\times D\times 1}$, Spach Transformer first obtained low-level feature embeddings ${\bf F}_0 \in \mathbb{R}^{H\times W\times D\times C}$ based on a $3\times 3\times 3$ convolution, where $H$, $W$, and $D$ indicate the voxel dimension and $C$ is the number of channels.  The four layers of the channel-wise encoder transformer blocks further processed the embedding  ${\bf F}_0$ to extract deep latent features  ${\bf F}_{l_c}$. 
	
	Additionally,  following \cite{dosovitskiy2020image,liu2021swin},  a patch partitioning module produced non-overlapping patches $\mathbf{x}_{i}\in \mathbb{R}^{P^3}$ from the low-dose PET image ${\bf X}$, where $P$ is the patch size and was set to 4 in our implementation.  A linear embedding layer was added to project these non-overlapping voxels to an arbitrary dimension as follows:
	\begin{equation} 
		\mathbf{z}_{0} =[\mathbf{x}_{0}\mathbf{E} ;\mathbf{x}_{1}\mathbf{E} ;\dotsc ;\mathbf{x}_{N_p}\mathbf{E}] \in \mathbb{R}^{N_{P} \times d},
	\end{equation}
	where $\mathbf{E} \in \mathbb{R}^{P^3 \times d}$ is a projection matrix, and 
	$\displaystyle N_{P} =\tfrac{H}{P} \times \tfrac{W}{P} \times \tfrac{D}{P}$ is the number of patches. 
	The spatial-wise encoder transformer blocks and the patch-merging layer used the tokens $\mathbf{z}_{0}$ to extract deep latent features  ${\bf F}_{l_s}$.  Role of the patch-merging layer is to reduce the spatial size by half along each dimension.  
	
	The encoder part described above gradually reduced the spatial size while expanding the channel capacity. The final latent features were generated by concatenating the latent features from the two transformer paths as  $ \mathbf{F}_{l} = \left[\mathbf{F}_{l_{c}} ,\mathbf{F}_{l_{s}}\right]$.  The final latent features $\mathbf{F}_{l}$ were further supplied to the decoder path to generate the high-quality PET image $\hat{\mathbf{X}}$.   Following \cite{zamir2021restormer},  a $1 \times 1 \times 1$ convolution and a concatenation operation was utilized to reduce the channel size by half at each decoder layer, followed by a channel-wise transformer block to recover the spatial information. The deep features  $\mathbf{F}_{d}$ were further refined to obtain robust features $\mathbf{F}_{r}$ at the PET image resolution scale.  Finally,  a $3 \times 3 \times 3$ convolution layer was applied on $\mathbf{F}_{r}$ to produce the residual image $\mathbf{R}$,  and the final denoised PET image was generated as $\hat{\mathbf{X}} = \mathbf{R} + \mathbf{X}$.
	
	\subsection{Spatial-wise Transformer Block} \label{sec.spatial}
	Swin Transformer \cite{liu2021swin} proposed to use window-based and shift window-based MSA modules to replace the standard MSA modules.  Fig. \ref{fig.sw_block} shows the diagram of the transformer block,  which is based on the following operations:
	\begin{equation}\label{eq.swin}
		\begin{aligned}
			\hat{\mathbf{z}}_{l} 	&=MSA_W\left(LN(\mathbf{z}_{l-1})\right) +\mathbf{z}_{l-1}, 		\\
			\mathbf{z}_{l} 			&=MLP\left(LN(\hat{\mathbf{z}}_{l})\right) +\hat{\mathbf{z}}_{l}, \\
			\hat{\mathbf{z}}_{l+1} 	&=MSA_{SW}\left(LN(\mathbf{z}_{l})\right) +\mathbf{z}_{l}, 			\\
			\mathbf{z}_{l+1} 		&=MLP\left(LN(\hat{\mathbf{z}}_{l+1})\right) +\hat{\mathbf{z}}_{l+1},
		\end{aligned}
	\end{equation}
	where $MSA_W$ and $MSA_{SW}$ are regular window-based MSA (W-MSA) and shifted window-based MSA (SW-MSA),  respectively,  $MLP$ is the multi-layer perceptron (MLP) module with the Gaussian error linear unit (GELU) \cite{hendrycks2016gaussian} as the activation layer,  and $LN$ is the layer normalization (LN) module \cite{ba2016layer} which was inserted before the W-MAS, SW-MSA and MLP modules.  Residual connection was used after each module.  Fig. ~\ref{fig.sw_sa} shows the window-based self-attention diagram.  The self-attention was calculated as  \cite{liu2021swin} :
	\begin{equation}
		\text{Attention}(\mathbf{Q}_s ,\mathbf{K}_s ,\mathbf{V}_s) =\text{Softmax}\left(\mathbf{Q}_s \mathbf{K}_s^{T} /\sqrt{d} +\mathbf{B}\right)\mathbf{V}_s,
	\end{equation}
	where $\mathbf{B}\in \mathbb{R}^{M^3 \times M^3}$ is the relative position of tokens in each window, $d$ is the query/key dimension. and $M^3$ is the number of patches in the 3D image window. The query $\mathbf{Q}_s$, key $\mathbf{K}_s$,  and value $\mathbf{V}_s$ matrices were computed as:
	\begin{equation}
		\mathbf{Q}_s =W_{l}^{Q}\mathbf{Y} ,\mathbf{K}_s =W_{l}^{K} \mathbf{Y} ,\mathbf{V}_s =W_{l}^{V}  \mathbf{Y},
	\end{equation}
	where $W_{l}^Q\in \mathbb{R}^{M^3 \times M^3}$, $W_{l}^K\in \mathbb{R}^{M^3 \times M^3}$ , and $W_{l}^V\in \mathbb{R}^{M^3 \times M^3}$ are the linear projection matrices,   and $\mathbf{Y} \in \mathbb{R}^{M^3 \times d}$ is the input after the LN layer.  
	\subsection{Channel-wise Transformer Block}\label{sec.channel}
	The Restormer work \cite{zamir2021restormer} designed the multi-Dconv head transposed attention (MDTA) modules to replace the standard MSA modules,  which computed self-attention along the channel dimension instead of the spatial dimension.  This can allow a global attention calculation with linearly computational complexity.  In Restormer,  each transformer block included a MDTA and a gated-Dconv feed-forward network (GDFN) module.  Though Restormer showed competing performance in natural image enhancement,  based on our experiments,  we found that this configuration has the pitfall of losing local spatial information.  Apart from transferring local spatial information in parallel using the spatial-wise transformer blocks as described in Sec.~\ref{sec.spatial},  in the proposed channel-wise transformer block (see Fig. \ref{fig.cw_block}), we further designed a gated-conv feed-forward network (GCFN) to better capture local spatial information and thus improve the lesion-uptake recovery.  Below we will introduce the details of MDTA, GDFN and GCFN. 
	\subsubsection{MDTA}
	Fig. \ref{fig.cw_mdta} shows the diagram of the MDTA module.  Given an input feature map ${\bf F} \in \mathbb{R}^{\hat{H} \times \hat{W} \times \hat{D} \times \hat{C}}$, a layer normalization module was first applied to obtain $\mathbf{Y} \in \mathbb{R}^{\hat{H} \times \hat{W} \times \hat{D} \times \hat{C}}$. The query $\mathbf{Q}_c  \in \mathbb{R}^{\hat{H} \times \hat{W} \times \hat{D} \times \hat{C}}$,  key $\mathbf{K}_c \in \mathbb{R}^{\hat{H} \times \hat{W} \times \hat{D} \times \hat{C}}$,  and value $\mathbf{V}_c \in \mathbb{R}^{\hat{H} \times \hat{W} \times \hat{D} \times \hat{C}}$ matrices were then obtained after a $1 \times 1 \times 1$ pixel-wise convolution operation (encoding channel-wise context) and a $3 \times 3 \times 3$ channel-wise convolution operation (aggregating pixel-wise cross-channel context).  By reshaping operations, $\hat{\mathbf{Q}}_c \in \mathbb{R}^{\hat{H}  \hat{W} \hat{D} \times \hat{C}}$, 
	$\hat{\mathbf{K}}_c \in \mathbb{R}^{\hat{C} \times \hat{H}  \hat{W} \hat{D}}$, and 
	$\hat{\mathbf{V}}_c \in \mathbb{R}^{\hat{H}  \hat{W} \hat{D} \times \hat{C}}$  were  obtained from  $\mathbf{Q}_c$, $\mathbf{K}_c$, and $\mathbf{V}_c$, respectively, and were used to generate a transposed-attention map $\mathbf{A} \in \mathbb{R}^{\hat{C} \times \hat{C}}$ instead of the massive spatial attention map size $\mathbb{R}^{  \hat{H}  \hat{W} \hat{D} \times  \hat{H}  \hat{W} \hat{D}}$ \cite{vaswani2017attention, dosovitskiy2020image}. The output feature of the MDTA module $\hat{\mathbf{F}}$ can be expressed as
	\begin{equation}
		\begin{aligned}
			& \hat{\mathbf{F}} =W_{p}\text{Attention}(\hat{\mathbf{Q}}_{c} ,\hat{\mathbf{K}}_{c} ,\hat{\mathbf{V}}_{c})+\mathbf{F},\\
			& \text{Attention}(\hat{\mathbf{Q}}_{c} ,\hat{\mathbf{K}}_{c} ,\hat{\mathbf{V}}_{c}) =\hat{\mathbf{V}}_{c} \cdot \text{Softmax}({\hat{\mathbf{Q}}_{c} \cdot }\hat{\mathbf{K}}_{c} / \alpha ),
		\end{aligned}
	\end{equation}
	where $\alpha$ is a learnable scaling parameter to control the magnitude of the dot product output ${\hat{\mathbf{Q}}_{c} \cdot }\hat{\mathbf{K}}_{c}$,  and $W_{p}$ is a linear projection matrix.  Note that the output of the self-attention operation was reshaped to have the same size as the input ${\bf F}$.

	\begin{table}[t] \centering
		\caption{According to channel number change $C$, an ablation study of different model parameters of Spach Transformer (GCFN).}
		\label{tbl.param_abl}
		\begin{tabular}{|@{\hskip3pt}c@{\hskip3pt}|c@{\hskip3pt}|@{\hskip3pt}r@{\hskip3pt}|cc|cc|}
			\hline
			{\rule{0pt}{3ex}}\multirow{2}{*}{}           & \multirow{2}{*}{C} & \multirow{2}{*}{\begin{tabular}[c]{@{}c@{}}Model\\ Parameters\end{tabular}} & \multicolumn{2}{c|}{PSNR}             & \multicolumn{2}{c|}{SSIM}            \\ \cline{4-7} 
			{\rule{0pt}{3ex}}&                    &                                                                             & \multicolumn{1}{c|}{Avg}     & Std    & \multicolumn{1}{c|}{Avg}    & Std    \\ \hline
			{\rule{0pt}{3ex}}\multirow{5}{*}{\begin{tabular}[c]{@{}c@{}}Spach \\ Transformer\end{tabular}} & 6                  & 7,034,951                                                                   & \multicolumn{1}{c|}{54.9753} & 3.6359 & \multicolumn{1}{c|}{0.9367} & 0.0147 \\ \cline{2-7} 
			{\rule{0pt}{3ex}}& 12                 & 19,218,323                                                                  & \multicolumn{1}{c|}{55.0840} & 3.6092 & \multicolumn{1}{c|}{\textbf{0.9403}} & \textbf{0.0128} \\ \cline{2-7} 
			{\rule{0pt}{3ex}}& 24                 & 59,350,043                                                                  & \multicolumn{1}{c|}{55.0911} & 3.6103 & \multicolumn{1}{c|}{0.9358} & 0.0147 \\ \cline{2-7} 
			{\rule{0pt}{3ex}}& 36                 & 120,501,731                                                                 & \multicolumn{1}{c|}{55.1003} & 3.6193 & \multicolumn{1}{c|}{0.9379} & 0.0137 \\ \cline{2-7} 
			{\rule{0pt}{3ex}}& 48                 & 202,673,387                                                                 & \multicolumn{1}{c|}{\textbf{55.1443}} & \textbf{3.6036} & \multicolumn{1}{c|}{0.9368} & 0.0138 \\ \hline
		\end{tabular}
	\end{table}
	\begin{figure*}[t]
		\centering
		\begin{subfigure}[b]{\textwidth}
			\centering
			\includegraphics[width=0.9\textwidth]{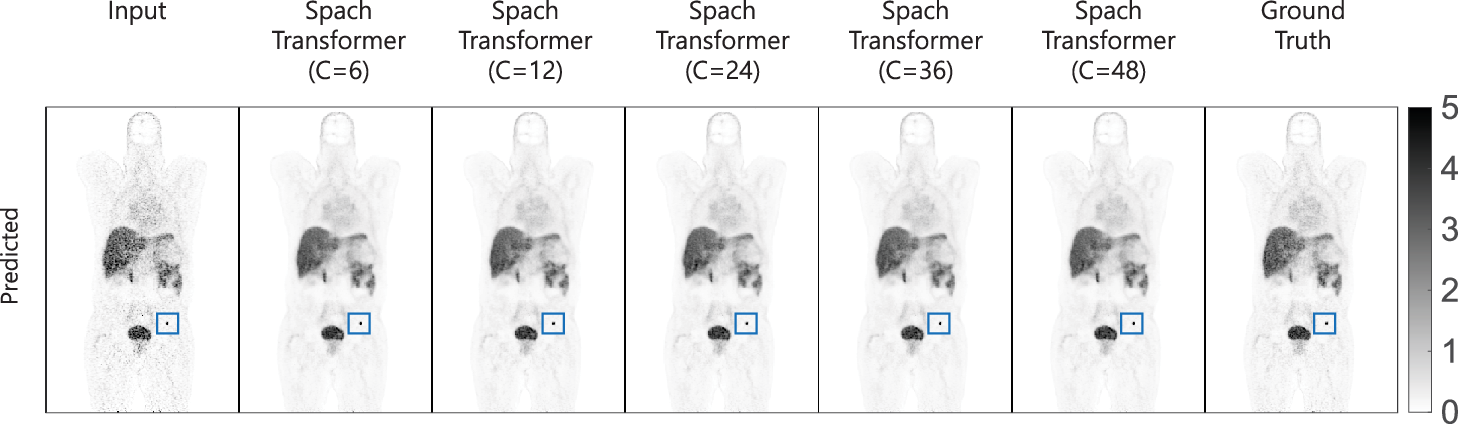}
			\caption{The predicted images and the distance images of the whole body.}
			\label{fig.abl1_1}
		\end{subfigure}
		\hfill
		\begin{subfigure}[b]{\textwidth}
			\centering
			\includegraphics[width=0.92\textwidth]{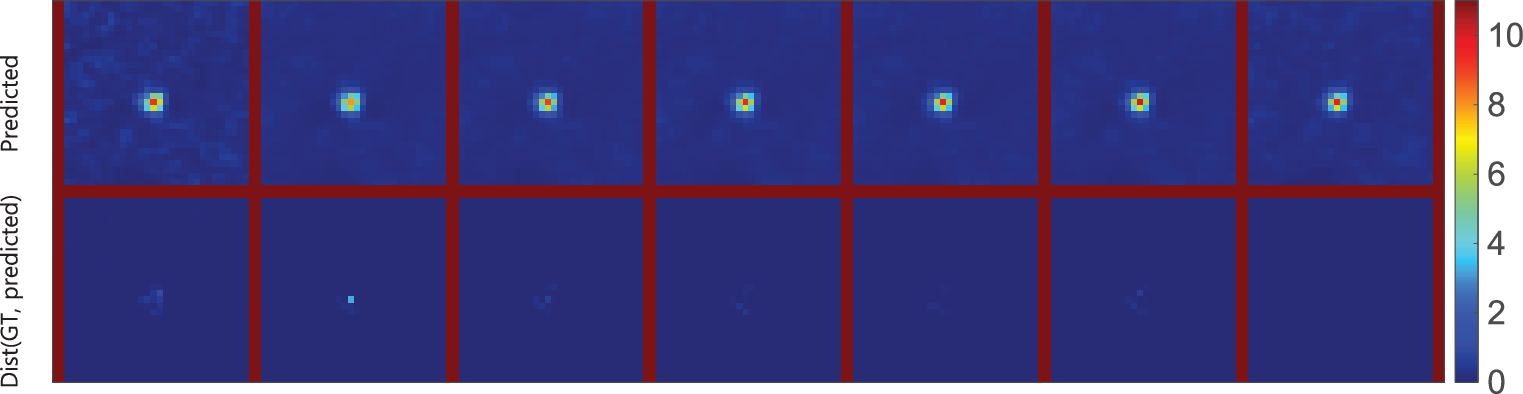}
			\caption{The enlarged images where the blue rectangulars are located in (a).}
			\label{fig.abl1_2}
		\end{subfigure}
		
		\caption{The denoised PET results of Spach Transformer according to different numbers of channels $C$ on an  $^{18}$F-DCFPyL dataset}
		\label{fig.abl1}
	\end{figure*}


	\subsubsection{GDFN}
	{\color{black}A gating mechanism \cite{dauphin2017language} aided in passing only helpful information to the next layer in the network architecture.
		Due to this, the gating network could suppress less informative features.
		In \cite{zamir2021restormer}, the gating mechanism was utilized in the GDFN module  (see Fig. \ref{fig.cw_gdfn}), which was calculated as}
	\begin{equation}
		\begin{aligned}
			& \hat{\mathbf{F}} =W_{p}^{0}\text{Gating}(\mathbf{F}) +\mathbf{F}, \\
			& \text{Gating}(\mathbf{F}) =\phi \left( W_{c}^{1} W_{p}^{1}{LN}(\mathbf{F})\right) \odot W_{c}^{2} W_{p}^{2}{LN}(\mathbf{F}).
		\end{aligned}
	\end{equation}
	Here ${\mathbf{F}}$ and $\hat{\mathbf{F}} $ are the input and output features, $\odot$ is an element-wise multiplication,  $\phi$ is the GELU non-linearity \cite{hendrycks2016gaussian},  and $LN$ indicates the LN layer. $W_p^0$, $W_p^1$, and $W_p^2$ indicate the linear projection matrices.  $W_{c}^{1}$ and $W_{c}^{1}$ represent  $3 \times 3 \times 3$ channel-wise convolutions.
	
	\subsubsection{GCFN}
	Since the above MDTA and GDFN modules focus on explicitly addressing channel information for image denoising, 
	it may cause losing spatial information.  To preserve spatial attributes between layers,  a gated-conv feed-forward network (GCFN) module (see Fig. \ref{fig.cw_gcfn}) was added to the channel-wise transformer block.  The only difference between GCFN and GDFN are that the $3 \times 3 \times 3$ channel-wise convolutions used in GDFN ($W_{c}^{1}$ and $W_{c}^{1}$) were replaced by $3 \times 3 \times 3$ spatial-wise convolutions ($W_{s}^{1}$ and $W_{s}^{1}$).  We hypothesized that the spatial-wise convolutions could better utilize spatial information and thus further help recovery lesion uptakes.

	\begin{figure*}[t!]
		\centering    
		\begin{subfigure}[h]{ \columnwidth}
			\centering
			\includegraphics[width=1\columnwidth]{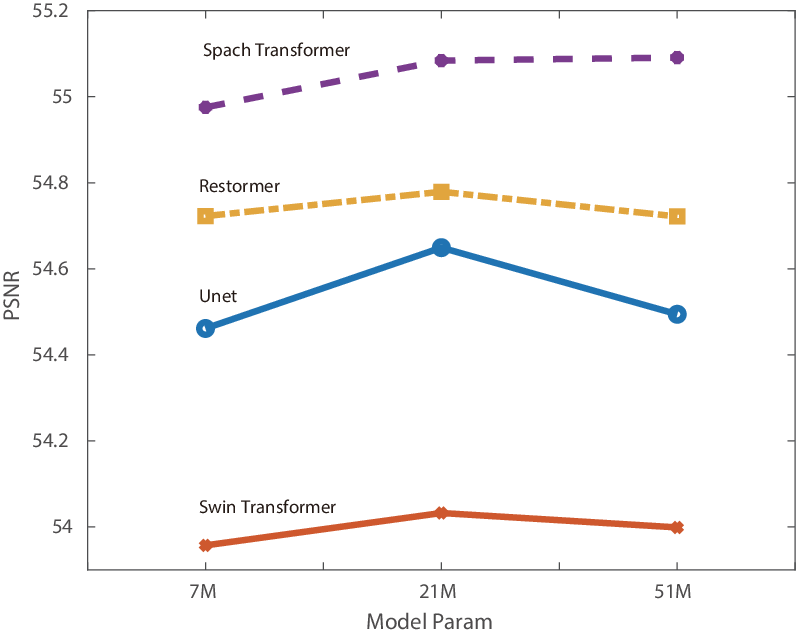} 
			\caption{PSNR} 
			\label{fig.PSNR_param} 
		\end{subfigure}   
		\begin{subfigure}[h]{ \columnwidth} 
			\centering
			\includegraphics[width=1\columnwidth]{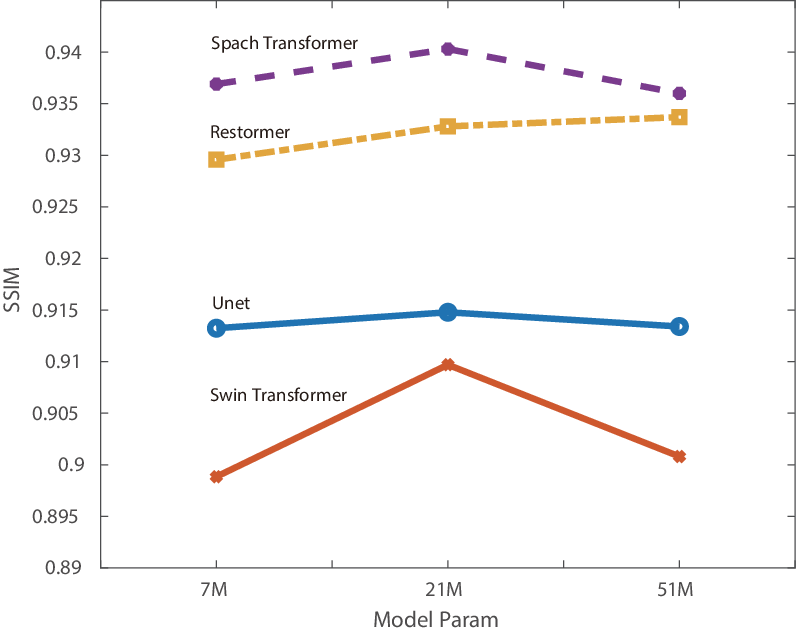}
			\caption{SSIM}
			\label{fig.SSIM_param}
		\end{subfigure} 
		
		\caption{Comparison of (a) PSNR and (b) SSIM for each method according to different model parameters.}
		\label{fig.param_study}
	\end{figure*}

	\begin{figure*}[t]
		\centering
		\begin{subfigure}[b]{\textwidth}
			\centering
			\includegraphics[width=0.9\textwidth]{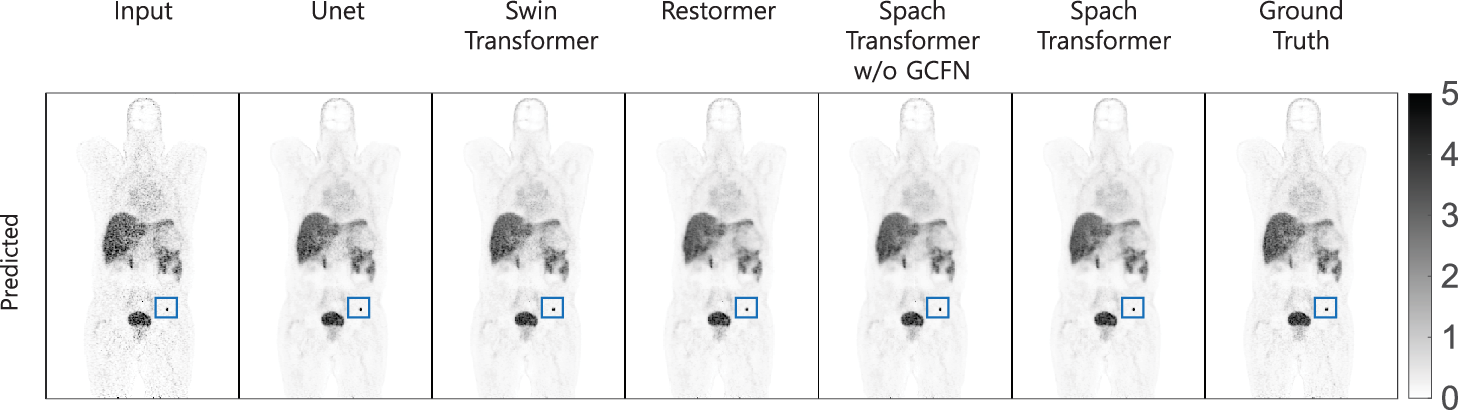}
			\caption{The predicted images and the distance images of the whole body.}
			\label{fig.exp1_1}
		\end{subfigure}
		\hfill
		\begin{subfigure}[b]{\textwidth}
			\centering
			\includegraphics[width=0.92\textwidth]{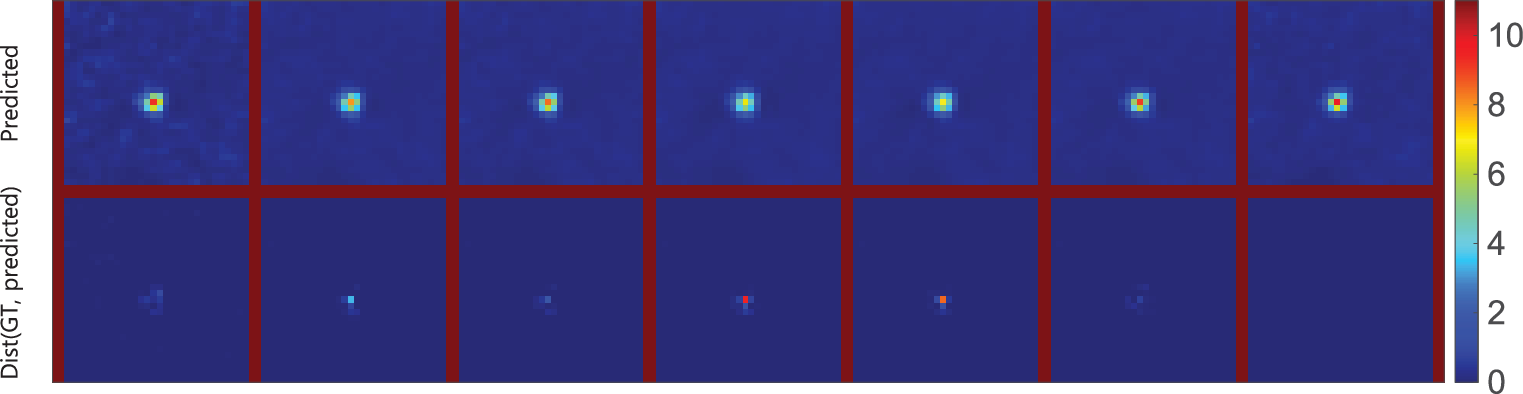}
			\caption{The enlarged images where the blue rectangulars are located in (a).}
			\label{fig.exp1_2}
		\end{subfigure}
		\caption{The denoised PET results of the state-of-the-art methods on a 1/4 low-dose  $^{18}$F-DCFPyL dataset.}
		\label{fig.exp1}
	\end{figure*}
	
	\section{Experiments}
	\subsection{Datasets}
	
	To evaluate the performance of the proposed Spach Transformer,  clinical whole-body $^{18}$F-FDG,  $^{18}$F-ACBC,  $^{18}$F-DCFPyL,  and $^{68}$Ga-DOTATATE datasets acquired from the GE DMI PET/CT scanner were utilized.  These tracers have unique distributions in the organs and are currently the most widely used tracers in oncology PET studies.  For each dataset,  1/4 and {\textcolor{myCOLOR}{1/16}} of the events were extracted from the listmode data to generate the 1/4 and {\textcolor{myCOLOR}{1/16}} low-dose data,  which was reconstructed and supplied as the network input.  {\textcolor{myCOLOR}{To investigate the consistent accuracy and precision of the methods, we carried out 10 iterations of list-mode data resampling for both the 1/4 and 1/16 low-dose datasets, utilizing different noise realizations each time.}} The corresponding normal-dose image was adopted as the training label.  Both the low-dose and normal-dose PET images were reconstructed based on the ordered subset expectation maximization (OSEM) algorithm (3 iterations and 17 subsets) with point spread function (PSF) and time of flight (TOF) modeling.  For the training set, 30  $^{18}$F-FDG and 30  $^{18}$F-ACBC datasets were included.  For the validation set,  4 $^{18}$F-FDG datasets were used.  For the testing set, 10  $^{18}$F-FDG, 10  $^{18}$F-ACBC,  10 $^{18}$F-DCFPyL,  and 18 $^{68}$Ga-DOTATATE datasets were employed.  Note that $^{18}$F-DCFPyL and  $^{68}$Ga-DOTATATE datasets were not included in the training set,  but in the testing set.  This is to evaluate the robustness of the deep learning structures to datasets of new tracers unseen during the training phase. The whole-body PET images had a dimension of $256 \times 256 \times 345$ and a voxel size of $2.73 \times 2.73 \times 2.8$ mm$^3$.  During network training, the input image was divided into $96 \times 96 \times 96$ cubes due to the GPU memory limit.  {\textcolor{myCOLOR}{The images were partially overlapped.}}

	\begin{figure*}[t]
		\centering
		\begin{subfigure}[b]{\textwidth}
			\centering
			\includegraphics[width=0.9\textwidth]{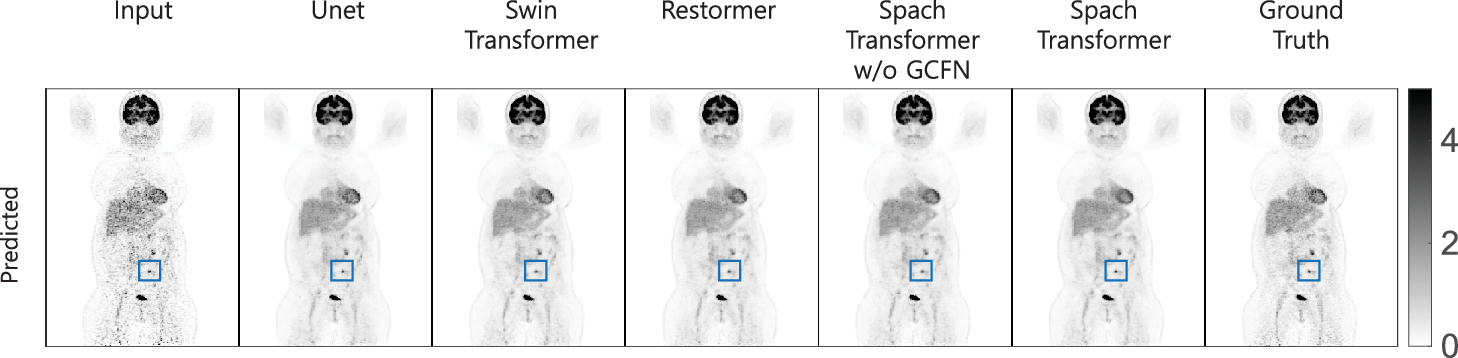}
			\caption{The predicted images and the distance images of the whole body.}
			\label{fig.exp2_1}
		\end{subfigure}
		\hfill
		\begin{subfigure}[b]{\textwidth}
			\centering
			\includegraphics[width=0.92\textwidth]{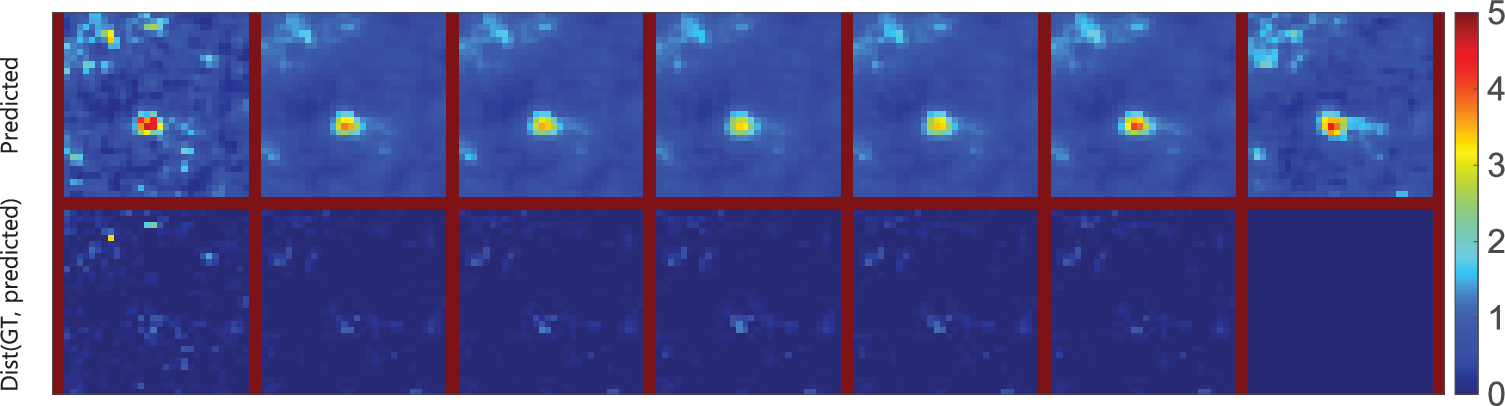}
			\caption{The enlarged images where the blue rectangulars are located in (a).}
			\label{fig.exp2_2}
		\end{subfigure} 
		\caption{The denoised PET results of the state-of-the-art methods on a 1/4 low-dose  $^{18}$F-FDG dataset.}
		\label{fig.exp2}
	\end{figure*}
	
	\subsection{Implementation and quantification}
	The Unet \cite{ronneberger2015u},  Swin Transformer \cite{liu2021swin, chen2021transmorph}, and Restormer \cite{zamir2021restormer} structures were adopted as the reference methods.
	\textcolor{myCOLOR}{For a fair comparison, all the structures were implemented with four levels, and the channel number $C$ was changed to have similar model parameters (21M). }
	All the architectures utilized the Charbonnier loss \cite{barron2019general} as the training loss function.  
	\textcolor{myCOLOR}{The best models of each architecture were chosen based on their best PSNR performances on the validation set.}
	For Spach Transformer,  the channel size of the transformer blocks can influence the network parameters a lot.  We have investigated the performance of the Spach Transformer with different channel sizes.  During network training, the AdamW optimizer was used and 300 epochs were run with the initial learning rate of 1e-5 gradually reduced to 1e-8 using cosine annealing \cite{loshchilov2016sgdr}.  The proposed network was implemented on Nvidia Tesla V100 GPU.
	
	As for quantification,  the peak signal-to-noise ratio (PSNR) and structural similarity index (SSIM) were {\textcolor{myCOLOR}{calculated using the whole-body PET images and}}  utilized to quantify the global performance of different methods.  For oncological PET images,  preserving tumor uptake is essential for disease diagnosis and monitoring.  To evaluate the performance of different methods regarding tumor-uptake recovery,  the contrast-to-noise (CNR) was calculated at the tumor region as
	\begin{equation}
		\text{CNR} = (V_{\text{tumor}} - V_{\text{ref}})/{\sigma}_{\text{ref}}.
	\end{equation}
	Here $V_{\text{tumor}}$ is the mean value of the tumor region of interest (ROI). \textcolor{myCOLOR}{The tumor ROI was selected based on the center of the tumor.} $V_{\text{ref}}$ and $\sigma_{\text{ref}}$ are the reference ROIs' mean value and standard deviation, respectively. The reference ROIs were selected from uniform regions inside the liver region.

	\section{Results}\label{sec.rst}
	\subsection{Effect of model parameters}\label{subsec.abl}
	{\color{black}We want to investigate the performance of Spach Transformer with different network parameters. The channel number used in the transformer architecture has a big impact on the model size and was varied to inverstigate the network performance with different network parameters.} Table~\ref{tbl.param_abl} shows the averaged PSNR and SSIM values for different model parameters of the Spach Transformer based on varying the channel size $C$ ($C=\{6, 12, 24, 36, 48\}$).  It can be observed that the best PSNR was achieved when $C=48$, while the best SSIM was obtained when $C=12$.  Fig. \ref{fig.abl1} shows the denoised PET images on one $^{18}$F-DCFPyL  dataset with different channel numbers. \textcolor{myCOLOR}{In all figures, such as Fig. \ref{fig.abl1}, the unit of the measurement of the color scale is consistently measured in standardized uptake value (SUV).}   To evaluate how close the predicted images are to the ground truth (GT), the $L^2$ norm distance images between the predicted and GT images were also computed.  It can be observed that when $C=6$,  the denoised image is blurred and uptake in the zoomed-in tumor region cannot be fully recovered.  When the channel number $C$ was increased to 12,  the network can better recover the tumor region and also has good denoising effect.  After further increasing the network parameter,  performance of the network on tumor-uptake recovery increases a little and becomes stable.  Based on the PSNR, SSIM results as well as the image appreance,  the channel size $C$ was set to 12 in all our studies. 
	\textcolor{myCOLOR}{In Fig. \ref{fig.param_study}, we also made a study for  the compared methods with different network parameters.
		Based on the parameter study, the comparison methods all were set to have network parameters of 21M, which is similar to the Spach Transformer with $C=12$ (19M).
	}

	\begin{figure*}[t]
		\centering
		\begin{subfigure}[b]{\textwidth}
			\centering
			\includegraphics[width=0.9\textwidth]{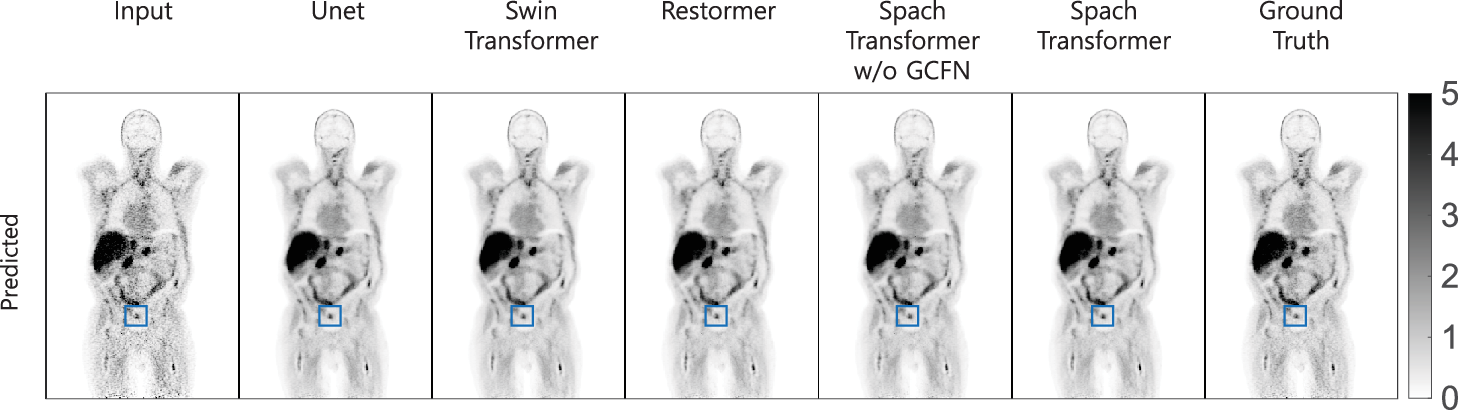}
			\caption{The predicted images and the distance images of the whole body.}
			\label{fig.exp3_1}
		\end{subfigure}
		\hfill
		\begin{subfigure}[b]{\textwidth}
			\centering
			\includegraphics[width=0.92\textwidth]{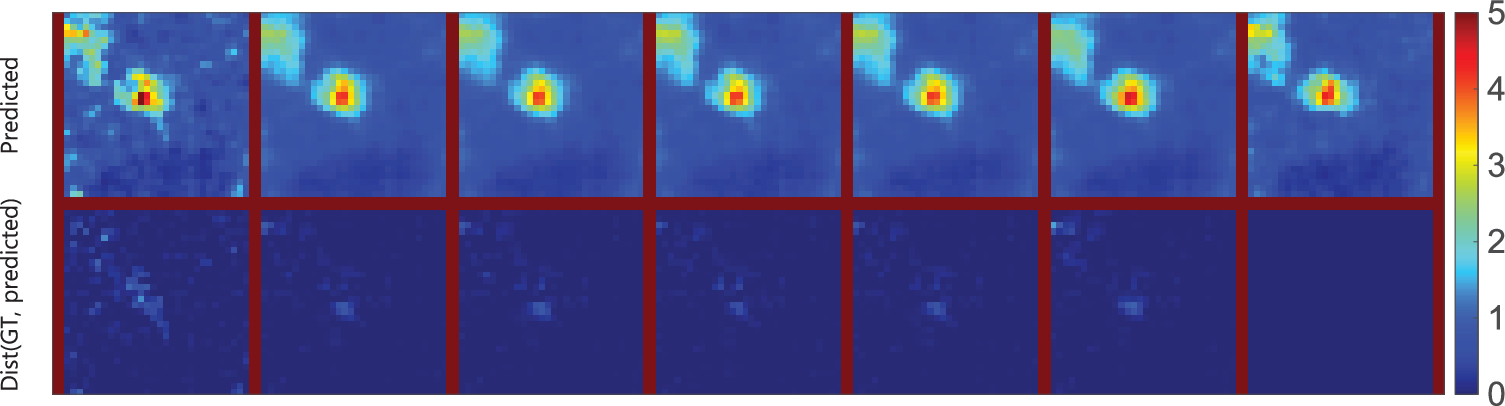}
			\caption{The enlarged images where the blue rectangulars are located in (a).}
			\label{fig.exp3_2}
		\end{subfigure}
		\caption{The denoised PET results of the state-of-the-art methods on a 1/4 low-dose  $^{18}$F-ACBC dataset.}
		\label{fig.exp3}
	\end{figure*}
	
	\begin{figure*}[t]
		\centering
		\begin{subfigure}[b]{\textwidth}
			\centering
			\includegraphics[width=0.9\textwidth]{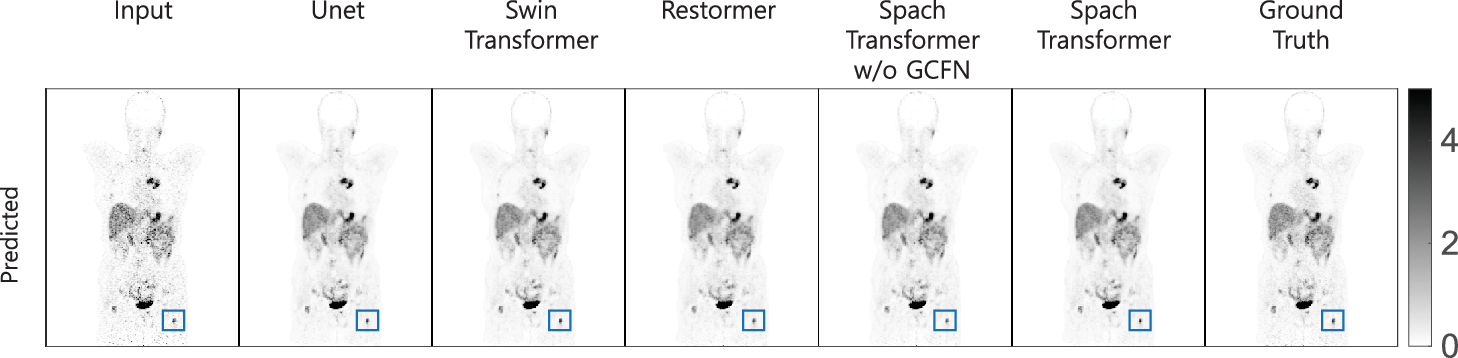}
			\caption{The predicted images and the distance images of the whole body.}
			\label{fig.exp4_1}
		\end{subfigure}
		\hfill
		\begin{subfigure}[b]{\textwidth}
			\centering
			\includegraphics[width=0.92\textwidth]{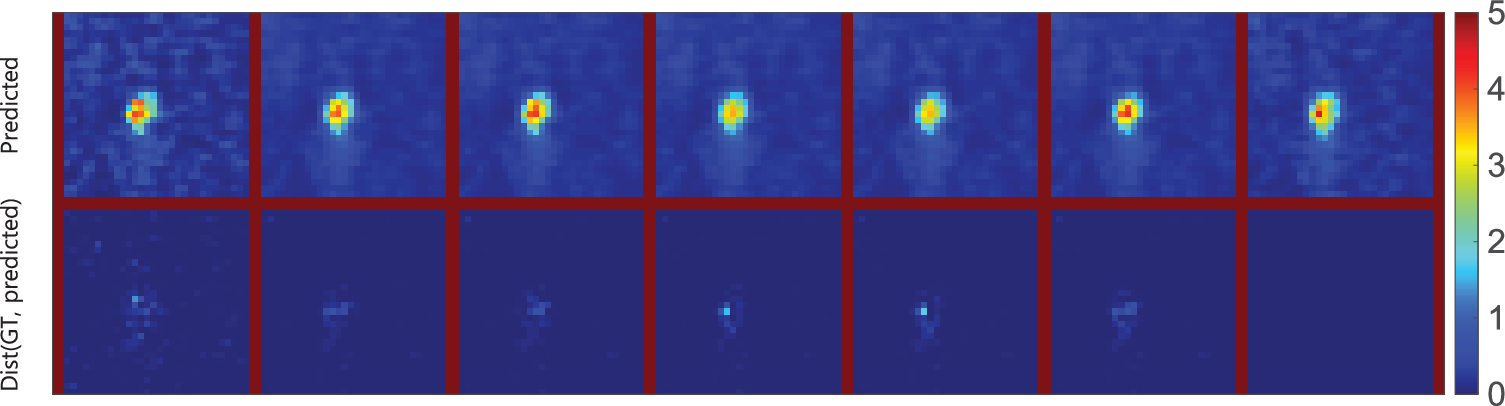}
			\caption{The enlarged images where the blue rectangulars are located in (a).}
			\label{fig.exp4_2}
		\end{subfigure}
		\caption{The denoised PET results of the state-of-the-art methods on a 1/4 low-dose Ga-68 DOTATATE dataset.}
		\label{fig.exp4}
	\end{figure*}

	\begin{table*}[t] \centering
		\caption{Quantitative comparison of PSNR with the state-of-the-art methods using the 1/4 low-dose datasets.}
		\label{tbl.psnr_sota}
		\begin{tabular}{|c|ccc|ccc|ccc|ccc|cc|}
			\hline
			{\rule{0pt}{3ex}}\multirow{3}{*}{Tracer}           & \multicolumn{3}{c|}{Unet}                                      & \multicolumn{3}{c|}{Swin Transformer}                          & \multicolumn{3}{c|}{Restormer}                                 & \multicolumn{3}{c|}{\begin{tabular}[c]{@{}c@{}}Spach Transformer\\ w/o GCFN\end{tabular}} & \multicolumn{2}{c|}{Spach Transformer}                          \\ \cline{2-15} 
			{\rule{0pt}{3ex}} & \multicolumn{1}{c|}{Mean}    & \multicolumn{1}{c|}{Std}    & H & \multicolumn{1}{c|}{Mean}    & \multicolumn{1}{c|}{Std}    & H & \multicolumn{1}{c|}{Mean}    & \multicolumn{1}{c|}{Std}    & H & \multicolumn{1}{c|}{Mean}           & \multicolumn{1}{c|}{Std}          & H       & \multicolumn{1}{c|}{Mean}             & Std             \\ \hline
			{\rule{0pt}{3ex}}$^{18}$F-DCFPyL  & \multicolumn{1}{c|}{53.9726} & \multicolumn{1}{c|}{5.7870} & 1 & \multicolumn{1}{c|}{53.5862} & \multicolumn{1}{c|}{5.7802} & 1 & \multicolumn{1}{c|}{53.8871} & \multicolumn{1}{c|}{5.7728} & 1 & \multicolumn{1}{c|}{53.9818}        & \multicolumn{1}{c|}{5.7754}       & 1       & \multicolumn{1}{c|}{\textbf{54.0927}} & \textbf{5.7879} \\ \hline
			{\rule{0pt}{3ex}}$^{18}$F-FDG     & \multicolumn{1}{c|}{52.6028} & \multicolumn{1}{c|}{3.1194} & 1 & \multicolumn{1}{c|}{52.1079} & \multicolumn{1}{c|}{3.2607} & 1 & \multicolumn{1}{c|}{52.5813} & \multicolumn{1}{c|}{3.0841} & 1 & \multicolumn{1}{c|}{52.6135}        & \multicolumn{1}{c|}{3.0513}       & 1       & \multicolumn{1}{c|}{\textbf{52.8678}} & \textbf{2.9585} \\ \hline
			{\rule{0pt}{3ex}}$^{18}$F-ACBC    & \multicolumn{1}{c|}{58.3996} & \multicolumn{1}{c|}{7.0715} & 1 & \multicolumn{1}{c|}{57.6852} & \multicolumn{1}{c|}{7.2052} & 1 & \multicolumn{1}{c|}{58.4845} & \multicolumn{1}{c|}{7.0278} & 1 & \multicolumn{1}{c|}{58.6009}        & \multicolumn{1}{c|}{7.0026}       & 1       & \multicolumn{1}{c|}{\textbf{58.7500}} & \textbf{6.9883} \\ \hline
			{\rule{0pt}{3ex}}Ga-68 DOTATATE   & \multicolumn{1}{c|}{54.0784} & \multicolumn{1}{c|}{3.7722} & 1 & \multicolumn{1}{c|}{53.3199} & \multicolumn{1}{c|}{3.8628} & 1 & \multicolumn{1}{c|}{54.4369} & \multicolumn{1}{c|}{3.7322} & 1 & \multicolumn{1}{c|}{54.4289}        & \multicolumn{1}{c|}{3.7438}       & 1       & \multicolumn{1}{c|}{\textbf{54.8294}} & \textbf{3.6679} \\ \hline
			{\rule{0pt}{3ex}}Avg.             & \multicolumn{1}{c|}{54.6492} & \multicolumn{1}{c|}{5.3646} & 1 & \multicolumn{1}{c|}{54.0323} & \multicolumn{1}{c|}{5.4194} & 1 & \multicolumn{1}{c|}{54.7790} & \multicolumn{1}{c|}{5.3398} & 1 & \multicolumn{1}{c|}{54.8267}        & \multicolumn{1}{c|}{5.3439}       & 1       & \multicolumn{1}{c|}{\textbf{55.0840}} & \textbf{5.2985} \\ \hline \hline
			{\rule{0pt}{3ex}}Model Parameters & \multicolumn{3}{c|}{21,781,520}                                & \multicolumn{3}{c|}{20,631,864}                                & \multicolumn{3}{c|}{21,156,805}                                & \multicolumn{3}{c|}{21,022,921}                                                   & \multicolumn{2}{c|}{19,218,323}                         \\ \hline
		\end{tabular}
	\end{table*}
	\begin{table*}[t] \centering
		\caption{Quantitative comparison of SSIM with the state-of-the-art methods using the 1/4 low-dose datasets.}
		\label{tbl.ssim_sota}
		\begin{tabular}{|c|ccc|ccc|ccc|ccc|cc|}
			\hline
			{\rule{0pt}{3ex}}\multirow{3}{*}{Tracer} & \multicolumn{3}{c|}{Unet}                                     & \multicolumn{3}{c|}{Swin Transformer}                         & \multicolumn{3}{c|}{Restormer}                                & \multicolumn{3}{c|}{\begin{tabular}[c]{@{}c@{}}Spach Transformer\\ w/o GCFN\end{tabular}} & \multicolumn{2}{c|}{Spach Transformer}                         \\ \cline{2-15} 
			{\rule{0pt}{3ex}}& \multicolumn{1}{c|}{Mean}   & \multicolumn{1}{c|}{Std}    & H & \multicolumn{1}{c|}{Mean}   & \multicolumn{1}{c|}{Std}    & H & \multicolumn{1}{c|}{Mean}   & \multicolumn{1}{c|}{Std}    & H & \multicolumn{1}{c|}{Mean}          & \multicolumn{1}{c|}{Std}           & H       & \multicolumn{1}{c|}{Mean}            & Std             \\ \hline
			{\rule{0pt}{3ex}}$^{18}$F-DCFPyL         & \multicolumn{1}{c|}{0.9234} & \multicolumn{1}{c|}{0.0076} & 1 & \multicolumn{1}{c|}{0.9125} & \multicolumn{1}{c|}{0.0080} & 1 & \multicolumn{1}{c|}{0.9370} & \multicolumn{1}{c|}{0.0078} & 1 & \multicolumn{1}{c|}{0.9369}        & \multicolumn{1}{c|}{0.0079}        & 1       & \multicolumn{1}{c|}{\textbf{0.9439}} & \textbf{0.0069} \\ \hline
			{\rule{0pt}{3ex}}$^{18}$F-FDG            & \multicolumn{1}{c|}{0.9364} & \multicolumn{1}{c|}{0.0063} & 1 & \multicolumn{1}{c|}{0.9282} & \multicolumn{1}{c|}{0.0079} & 1 & \multicolumn{1}{c|}{0.9472} & \multicolumn{1}{c|}{0.0075} & 1 & \multicolumn{1}{c|}{0.9477}        & \multicolumn{1}{c|}{0.0076}        & 1       & \multicolumn{1}{c|}{\textbf{0.9539}} & \textbf{0.0074} \\ \hline
			{\rule{0pt}{3ex}}$^{18}$F-ACBC           & \multicolumn{1}{c|}{0.8894} & \multicolumn{1}{c|}{0.0059} & 1 & \multicolumn{1}{c|}{0.8759} & \multicolumn{1}{c|}{0.0052} & 1 & \multicolumn{1}{c|}{0.8955} & \multicolumn{1}{c|}{0.0061} & 1 & \multicolumn{1}{c|}{0.8975}        & \multicolumn{1}{c|}{0.0059}        & 1       & \multicolumn{1}{c|}{\textbf{0.9138}} & \textbf{0.0048} \\ \hline
			{\rule{0pt}{3ex}}Ga-68 DOTATATE          & \multicolumn{1}{c|}{0.9121} & \multicolumn{1}{c|}{0.0168} & 1 & \multicolumn{1}{c|}{0.9166} & \multicolumn{1}{c|}{0.0122} & 1 & \multicolumn{1}{c|}{0.9432} & \multicolumn{1}{c|}{0.0119} & 1 & \multicolumn{1}{c|}{0.9417}        & \multicolumn{1}{c|}{0.0129}        & 1       & \multicolumn{1}{c|}{\textbf{0.9458}} & \textbf{0.0127} \\ \hline
			{\rule{0pt}{3ex}}Avg.                    & \multicolumn{1}{c|}{0.9148} & \multicolumn{1}{c|}{0.0196} & 1 & \multicolumn{1}{c|}{0.9097} & \multicolumn{1}{c|}{0.0204} & 1 & \multicolumn{1}{c|}{0.9328} & \multicolumn{1}{c|}{0.0215} & 1 & \multicolumn{1}{c|}{0.9327}        & \multicolumn{1}{c|}{0.0208}        & 1       & \multicolumn{1}{c|}{\textbf{0.9404}} & \textbf{0.0169} \\ \hline
		\end{tabular}
	\end{table*}

	\begin{table*}[t] \centering
		\caption{Quantitative comparison of PSNR with the state-of-the-art methods using the 1/16 low-dose datasets.}
		\label{tbl.psnr_sota_16}
		\begin{tabular}{|c|ccc|ccc|ccc|ccc|cc|}
			\hline
			{\rule{0pt}{3ex}}\multirow{3}{*}{Tracer} & \multicolumn{3}{c|}{Unet}                                     & \multicolumn{3}{c|}{Swin Transformer}                         & \multicolumn{3}{c|}{Restormer}                                & \multicolumn{3}{c|}{\begin{tabular}[c]{@{}c@{}}Spach Transformer\\ w/o GCFN\end{tabular}} & \multicolumn{2}{c|}{Spach Transformer}                         \\ \cline{2-15} 
			{\rule{0pt}{3ex}}& \multicolumn{1}{c|}{Mean}   & \multicolumn{1}{c|}{Std}    & H & \multicolumn{1}{c|}{Mean}   & \multicolumn{1}{c|}{Std}    & H & \multicolumn{1}{c|}{Mean}   & \multicolumn{1}{c|}{Std}    & H & \multicolumn{1}{c|}{Mean}          & \multicolumn{1}{c|}{Std}           & H       & \multicolumn{1}{c|}{Mean}            & Std             \\ \hline
			{\rule{0pt}{3ex}}$^{18}$F-DCFPyL  & \multicolumn{1}{c|}{53.2089} & \multicolumn{1}{c|}{5.3284} & 1 & \multicolumn{1}{c|}{53.1360} & \multicolumn{1}{c|}{5.5652} & 1 & \multicolumn{1}{c|}{54.8250} & \multicolumn{1}{c|}{5.6157} & 1 & \multicolumn{1}{c|}{54.8070}          & \multicolumn{1}{c|}{5.5878}          & 1          & \multicolumn{1}{c|}{\textbf{55.3344}} & \textbf{5.5678} \\ \hline
			{\rule{0pt}{3ex}}$^{18}$F-FDG     & \multicolumn{1}{c|}{60.6127} & \multicolumn{1}{c|}{3.8768} & 1 & \multicolumn{1}{c|}{59.1802} & \multicolumn{1}{c|}{4.0072} & 1 & \multicolumn{1}{c|}{60.4522} & \multicolumn{1}{c|}{3.8358} & 1 & \multicolumn{1}{c|}{60.5567}          & \multicolumn{1}{c|}{3.8055}          & 1          & \multicolumn{1}{c|}{\textbf{60.8443}} & \textbf{3.8462} \\ \hline
			{\rule{0pt}{3ex}}$^{18}$F-ACBC    & \multicolumn{1}{c|}{50.8125} & \multicolumn{1}{c|}{5.2460} & 1 & \multicolumn{1}{c|}{49.2243} & \multicolumn{1}{c|}{5.7844} & 1 & \multicolumn{1}{c|}{50.5470} & \multicolumn{1}{c|}{5.2809} & 1 & \multicolumn{1}{c|}{50.5847}          & \multicolumn{1}{c|}{5.3249}          & 1          & \multicolumn{1}{c|}{\textbf{50.9787}} & \textbf{5.2225} \\ \hline
			{\rule{0pt}{3ex}}Ga-68 DOTATATE   & \multicolumn{1}{c|}{47.6882} & \multicolumn{1}{c|}{3.8545} & 1 & \multicolumn{1}{c|}{46.2906} & \multicolumn{1}{c|}{3.7951} & 1 & \multicolumn{1}{c|}{47.5465} & \multicolumn{1}{c|}{3.9095} & 1 & \multicolumn{1}{c|}{47.6477}          & \multicolumn{1}{c|}{4.0441}          & 1          & \multicolumn{1}{c|}{\textbf{48.0439}} & \textbf{4.1956} \\ \hline
			{\rule{0pt}{3ex}}Avg.             & \multicolumn{1}{c|}{53.1824} & \multicolumn{1}{c|}{5.9643} & 1 & \multicolumn{1}{c|}{52.2018} & \multicolumn{1}{c|}{6.1312} & 1 & \multicolumn{1}{c|}{53.5978} & \multicolumn{1}{c|}{6.1121} & 1 & \multicolumn{1}{c|}{53.6511}          & \multicolumn{1}{c|}{6.1167}          & 1          & \multicolumn{1}{c|}{\textbf{54.0678}} & \textbf{6.1032} \\ \hline
		\end{tabular}
	\end{table*}
	
	\begin{table*}[t] \centering
		\caption{Quantitative comparison of SSIM with the state-of-the-art methods using the 1/16 low-dose datasets.}
		\label{tbl.ssim_sota_16}
		\begin{tabular}{|c|ccc|ccc|ccc|ccc|cc|}
			\hline
			{\rule{0pt}{3ex}}\multirow{3}{*}{Tracer} & \multicolumn{3}{c|}{Unet}                                     & \multicolumn{3}{c|}{Swin Transformer}                         & \multicolumn{3}{c|}{Restormer}                                & \multicolumn{3}{c|}{\begin{tabular}[c]{@{}c@{}}Spach Transformer\\ w/o GCFN\end{tabular}} & \multicolumn{2}{c|}{Spach Transformer}                         \\ \cline{2-15} 
			{\rule{0pt}{3ex}}& \multicolumn{1}{c|}{Mean}   & \multicolumn{1}{c|}{Std}    & H & \multicolumn{1}{c|}{Mean}   & \multicolumn{1}{c|}{Std}    & H & \multicolumn{1}{c|}{Mean}   & \multicolumn{1}{c|}{Std}    & H & \multicolumn{1}{c|}{Mean}          & \multicolumn{1}{c|}{Std}           & H       & \multicolumn{1}{c|}{Mean}            & Std             \\ \hline
			{\rule{0pt}{3ex}}$^{18}$F-DCFPyL         & \multicolumn{1}{c|}{0.8755} & \multicolumn{1}{c|}{0.0461} & 1 & \multicolumn{1}{c|}{0.8634} & \multicolumn{1}{c|}{0.0351} & 1 & \multicolumn{1}{c|}{0.9108} & \multicolumn{1}{c|}{0.0213} & 1 & \multicolumn{1}{c|}{0.9099}           & \multicolumn{1}{c|}{0.0210}          & 1          & \multicolumn{1}{c|}{\textbf{0.9144}} & \textbf{0.0203} \\ \hline
			{\rule{0pt}{3ex}}$^{18}$F-FDG            & \multicolumn{1}{c|}{0.8808} & \multicolumn{1}{c|}{0.0052} & 1 & \multicolumn{1}{c|}{0.8600} & \multicolumn{1}{c|}{0.0039} & 1 & \multicolumn{1}{c|}{0.8803} & \multicolumn{1}{c|}{0.0070} & 1 & \multicolumn{1}{c|}{0.8775}           & \multicolumn{1}{c|}{0.0064}          & 1          & \multicolumn{1}{c|}{\textbf{0.8826}} & \textbf{0.0073} \\ \hline
			{\rule{0pt}{3ex}}$^{18}$F-ACBC           & \multicolumn{1}{c|}{0.9419} & \multicolumn{1}{c|}{0.0117} & 1 & \multicolumn{1}{c|}{0.9188} & \multicolumn{1}{c|}{0.0145} & 1 & \multicolumn{1}{c|}{0.9403} & \multicolumn{1}{c|}{0.0114} & 1 & \multicolumn{1}{c|}{0.9390}           & \multicolumn{1}{c|}{0.0112}          & 1          & \multicolumn{1}{c|}{\textbf{0.9425}} & \textbf{0.0109} \\ \hline
			{\rule{0pt}{3ex}}Ga-68 DOTATATE          & \multicolumn{1}{c|}{0.9252} & \multicolumn{1}{c|}{0.0147} & 1 & \multicolumn{1}{c|}{0.8918} & \multicolumn{1}{c|}{0.0251} & 1 & \multicolumn{1}{c|}{0.9267} & \multicolumn{1}{c|}{0.0127} & 1 & \multicolumn{1}{c|}{0.9249}           & \multicolumn{1}{c|}{0.0126}          & 1          & \multicolumn{1}{c|}{\textbf{0.9290}} & \textbf{0.0121} \\ \hline
			{\rule{0pt}{3ex}}Avg.                    & \multicolumn{1}{c|}{0.9031} & \multicolumn{1}{c|}{0.0391} & 1 & \multicolumn{1}{c|}{0.8812} & \multicolumn{1}{c|}{0.0327} & 1 & \multicolumn{1}{c|}{0.9149} & \multicolumn{1}{c|}{0.0254} & 1 & \multicolumn{1}{c|}{0.9134}           & \multicolumn{1}{c|}{0.0257}          & 1          & \multicolumn{1}{c|}{\textbf{0.9176}} & \textbf{0.0249} \\ \hline
		\end{tabular}
	\end{table*}

	\begin{figure*}[t!]
		\centering   \hspace{-0.2cm}
		\begin{subfigure}[h]{0.5\textwidth}
			\centering
			\includegraphics[width=0.92\textwidth]{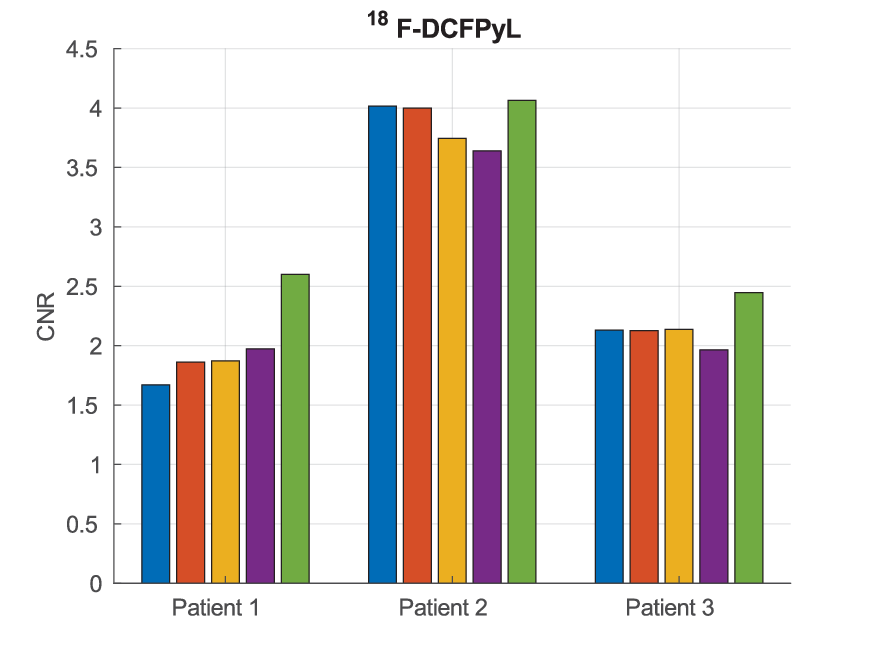}
			\caption{$^{18}$F-DCFPyL} 
			\label{fig.CNR1} 
		\end{subfigure}  \hspace{-0.2cm}
		\begin{subfigure}[h]{0.5\textwidth} 
			\centering
			\includegraphics[width=0.92\textwidth]{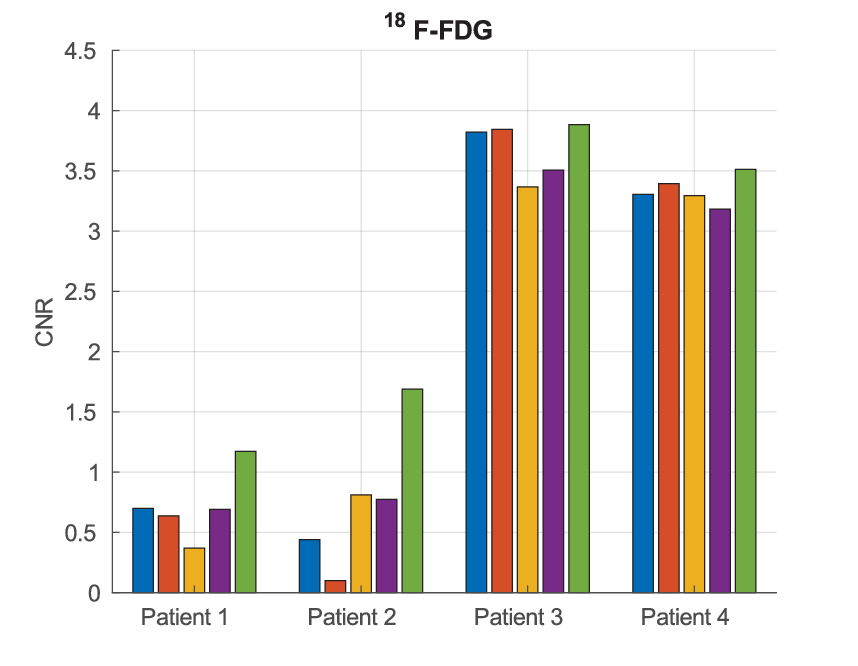}
			\caption{$^{18}$F-FDG}
			\label{fig.CNR2}
		\end{subfigure} 
		\hfill
		\hspace{-2.5cm}
		\begin{subfigure}[b]{0.5\textwidth}  
			\centering
			\includegraphics[width=0.92\textwidth]{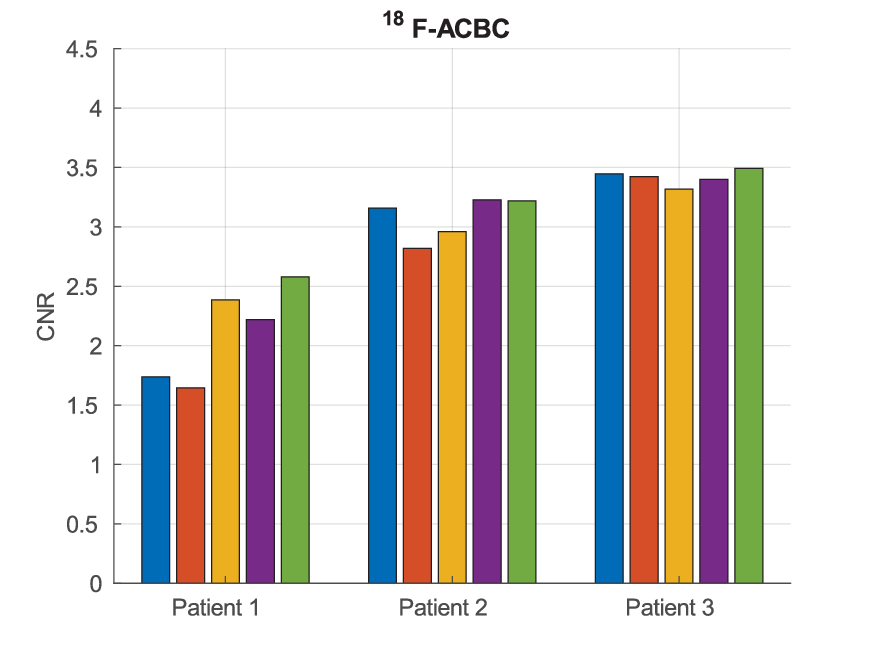}
			\caption{$^{18}$F-ACBC}
			\label{fig.CNR3}
		\end{subfigure} \hspace{-0.3cm}
		\begin{subfigure}[b]{0.5\textwidth}
			\centering
			\includegraphics[width=0.92\textwidth]{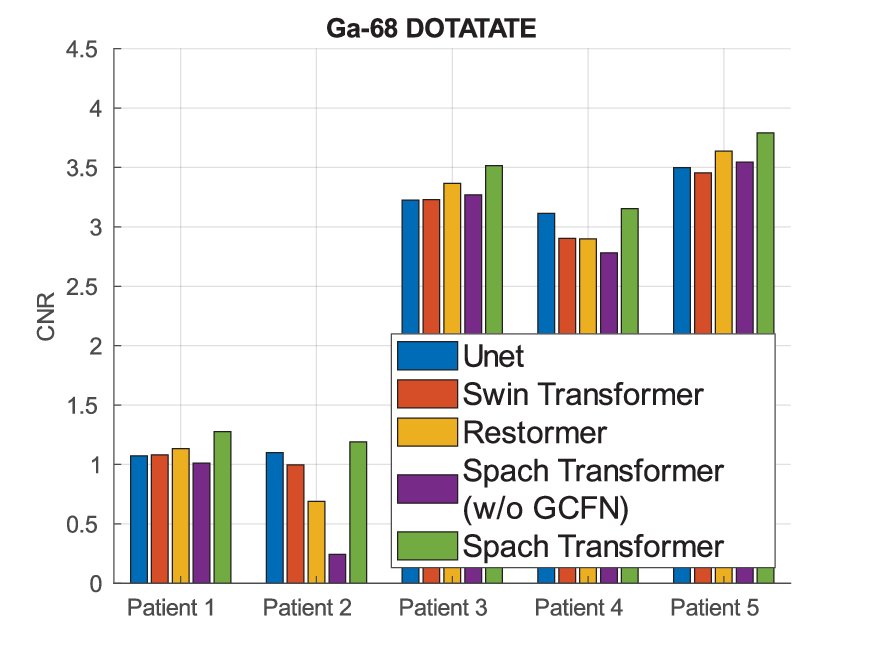}
			\caption{Ga-68 DOTATATE}
			\label{fig.CNR4}
		\end{subfigure} 
		\caption{Comparison of CNR with different state-of-the-art methods on the 1/4 low-dose (a) $^{18}$F-DCFPyL, (b) $^{18}$F-FDG, (c) $^{18}$F-ACBC and (d) Ga-68 DOTATATE  datasets.}
		\label{fig.CNR}
	\end{figure*}
	\begin{figure}[!t] 
		\centerline{\includegraphics[width=0.8\columnwidth]{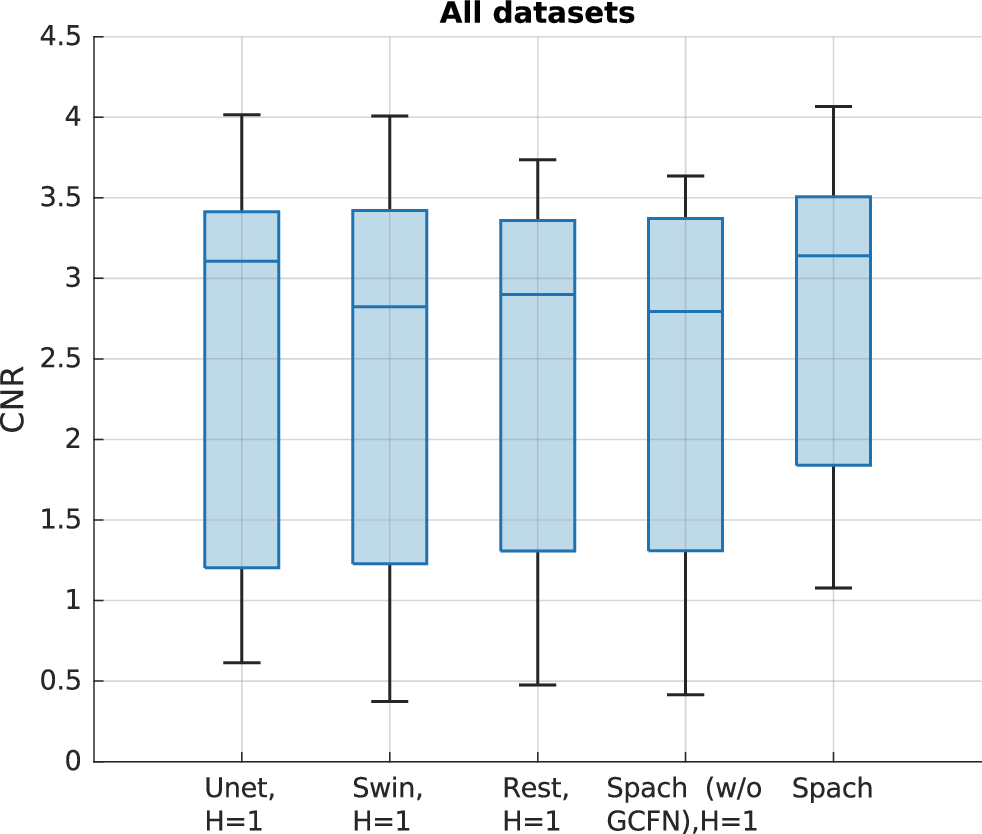}}
		\caption{A box plot of CNR for the 1/4 low-dose datasets shown in Fig.~\ref{fig.CNR}. In the legend for each method, `1' was given when the null hypothesis (H) was rejected at a significance level of 0.05; otherwise `0'.}
		\label{fig.cnr_box}
	\end{figure}
	\begin{figure*}[t!]
		\centering   \hspace{-0.2cm}
		\begin{subfigure}[h]{0.5\textwidth}
			\centering
			\includegraphics[width=0.92\textwidth]{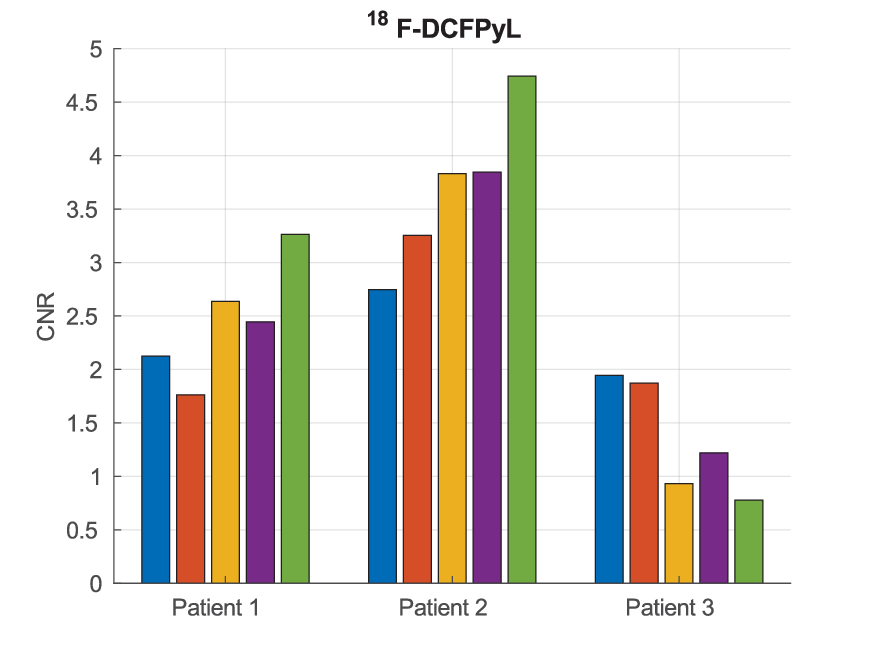}
			\caption{$^{18}$F-DCFPyL} 
			\label{fig.CNR_low_1} 
		\end{subfigure}  \hspace{-0.2cm}
		\begin{subfigure}[h]{0.5\textwidth} 
			\centering
			\includegraphics[width=0.92\textwidth]{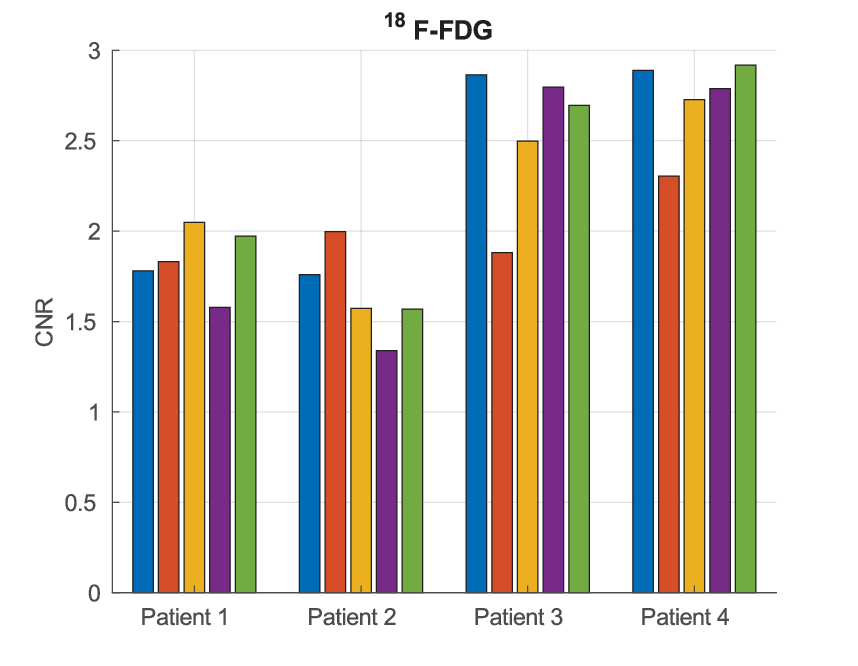}
			\caption{$^{18}$F-FDG}
			\label{fig.CNR_low_2}
		\end{subfigure} 
		\hfill
		\hspace{-2.5cm}
		\begin{subfigure}[b]{0.5\textwidth}  
			\centering
			\includegraphics[width=0.92\textwidth]{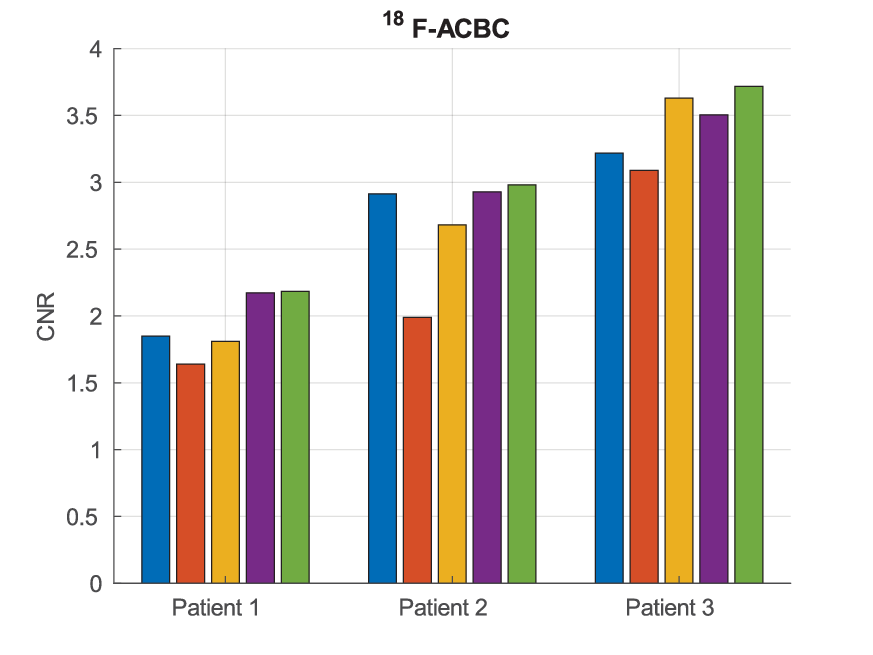}
			\caption{$^{18}$F-ACBC}
			\label{fig.CNR_low_3}
		\end{subfigure} \hspace{-0.3cm}
		\begin{subfigure}[b]{0.5\textwidth}
			\centering
			\includegraphics[width=0.92\textwidth]{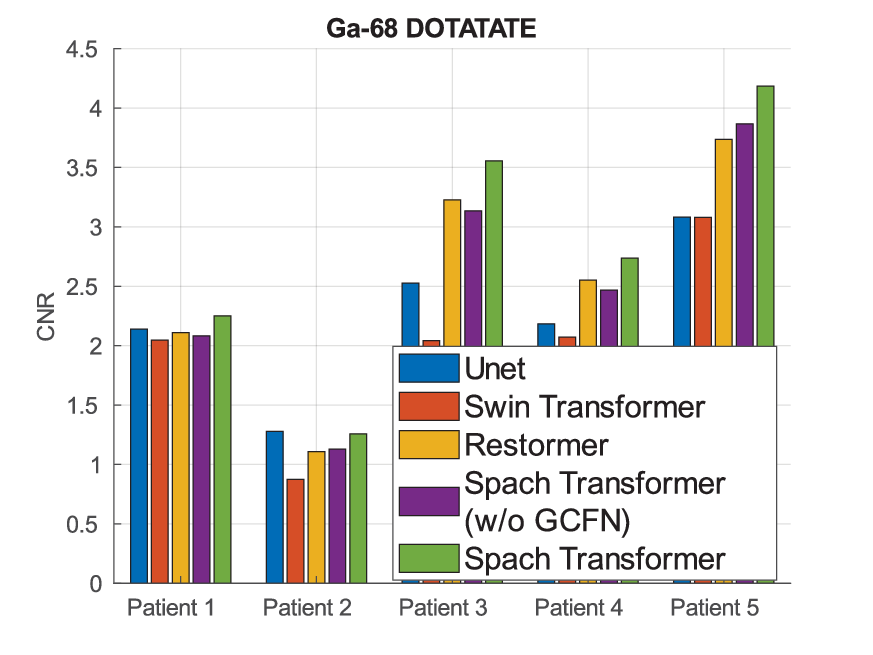}
			\caption{Ga-68 DOTATATE}
			\label{fig.CNR_low_4}
		\end{subfigure} 
		\caption{Comparison of CNR with different state-of-the-art methods on the 1/16 low-dose (a) $^{18}$F-DCFPyL, (b) $^{18}$F-FDG, (c) $^{18}$F-ACBC and (d) Ga-68 DOTATATE datasets.}
		\label{fig.CNR_low}
	\end{figure*}
	\begin{figure}[!t] 
		\centerline{\includegraphics[width=0.9\columnwidth]{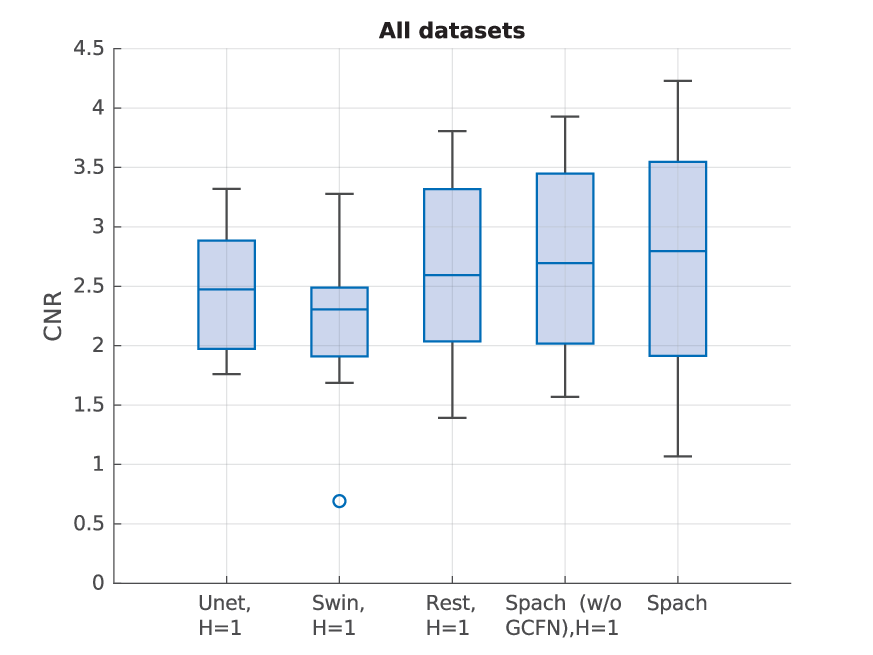}}
		\caption{A box plot of CNR for the 1/16 low-dose datasets shown in Fig.~\ref{fig.CNR_low}. In the legend for each method, `1' was given when the null hypothesis (H) was rejected at a significance level of 0.05; otherwise `0'.}
		\label{fig.cnr_box_low}
	\end{figure}

	\subsection{Comparisons with reference methods}
	Fig. \ref{fig.exp1}, \ref{fig.exp2}, \ref{fig.exp3}, and \ref{fig.exp4} show the results of the Spach Transformer and the reference methods based on 1/4 low-dose $^{18}$F-DCFPyL, $^{18}$F-FDG,  $^{18}$F-ACBC, and Ga-68 DOTATATE datasets. 
	In Fig. \ref{fig.exp1_1}, \ref{fig.exp2_1}, \ref{fig.exp3_1}, and \ref{fig.exp4_1}, the Restormer can obtain better denoising performances than the Unet and Swin Transformer from the input low-dose images.
	In  Fig. \ref{fig.exp1_2}, \ref{fig.exp2_2}, \ref{fig.exp3_2}, and \ref{fig.exp4_2}, the Unet and Swin Transformer can achieve relatively higher uptake values than the Restormer in the tumor regions. 
	This observation is from that the Unet and Swin Transformer focused on spatial information, whereas the Restormer was devoted to channel information.
	Since the Spach Transformer can process spatial and channel information together, our architecture achieved higher uptake values and better denoising performances than the other reference methods.
	\textcolor{myCOLOR}{Similar to Fig. \ref{fig.exp1}, \ref{fig.exp2}, \ref{fig.exp3}, and \ref{fig.exp4}, the results of the Spach Transformer and the reference methods, based on the 1/16 low-dose datasets, have been provided in the supplementary material.}
	
	
	In Table \ref{tbl.psnr_sota} and \ref{tbl.ssim_sota}, the Spach Transformer  achieved average PSNR of 55.0840 and average SSIM of 0.9404 on the 1/4 low-dose datasets. 
	In terms of the PSNR and SSIM, the Spach Transformer (19M) outperformed the Unet (21M), Swin Transformer (20M), Restormer (21M), and Spach Transformer without GCFN (21M).
	\textcolor{myCOLOR}{In Table \ref{tbl.psnr_sota} and \ref{tbl.ssim_sota}, we performed a statistical $t$-test, to check whether the proposed method significantly outperformed the compared methods. `1' showed that the null hypothesis (H) was rejected at a significance level of 0.05, otherwise, `0'. The proposed method significantly outperformed all the compared methods in each tracer and all the tracers.}
	Compared to the recent transformer model, the Restormer \cite{zamir2021restormer}, the Spach Transformer achieved a PSNR improvement of 0.305 and an SSIM performance gain of 0.0076.
	Fig. \ref{fig.CNR} shows the CNR results of the reference methods on the 1/4 low-dose datasets for each patient.
	\textcolor{myCOLOR}{Fig. \ref{fig.cnr_box} shows a box plot of the CNR results with the $t$-test. The horizontal lines in each box are median.}
	The Spach Transformer achieved the highest CNR results among all four tracers in terms of the box plot.

	\textcolor{myCOLOR}{For the 1/16 low-dose datasets, Table \ref{tbl.psnr_sota_16} and Table \ref{tbl.ssim_sota_16} also presented the averaged PSNR and SSIM values. The Spach Transformer achieved an average PSNR of 54.0678 and an average SSIM of 0.9176. Compared to the Restormer, the Spach Transformer demonstrated a PSNR improvement of 0.47 and an SSIM performance gain of 0.0027. Fig. \ref{fig.CNR_low} illustrates the CNR results of the reference methods for each patient, while Fig. \ref{fig.cnr_box_low} shows a box plot of the CNR results with the $t$-test. On the 1/16 low-dose datasets, the Spach Transformer attained the highest CNR results among all four tracers in terms of the box plot.}
	
	\subsection{Effect of GCFN}
	We conducted comparisons to examine the effectiveness of GCFN on the channel-wise transformer block of the Spach Transformer.
	Table \ref{tbl.psnr_sota} and \ref{tbl.ssim_sota} include an ablation study to check the effectiveness of GCFN quantitatively.
	The proposed GCFN achieved PSNR and SSIM performance gains on the Spach Transformer. 
	In order to investigate further, Fig. \ref{fig.CNR} illustrates each tracer's CNR results of the reference methods.
	\textcolor{myCOLOR}{The GCFN helped the proposed method to significantly improve the CNR performances compared to all the compared methods.}
	Fig. \ref{fig.exp1}, \ref{fig.exp2}, \ref{fig.exp3}, and \ref{fig.exp4} also include an ablation study of denoised PET results.
	The GCFN helped to obtain higher uptake values than the exclusion of GCFN in the tumor regions.

	\section{Discussion}
	For PET image denoising, it is hard to reduce the noise from the low-dose injection and simultaneously improve the tumor region's uptake values. 
	The latest deep learning architectures (e.g., the Swin Transformer and the Restormer) have focused on spatial or channel information.
	\textcolor{myCOLOR}{In this work, we proposed a novel spatial and channel-wise transformer architecture, Spach Transformer, to suppress the noise and highlight the uptake values at the same time.}
	In the Section \ref{sec.rst}, we observed that the spatial information-oriented network architectures (e.g., the Unet and Swin Transformer) could highlight the tumor region well but were still noisy.
	In contrast, the channel information-oriented network architecture (e.g., the Restormer) could achieve on better denoising performances than the other reference methods but failed to find high uptake values.
	We could see such observations through the reported quantitative and qualitative results.
	Table \ref{tbl.psnr_sota}, \ref{tbl.ssim_sota}, \ref{tbl.psnr_sota_16},   and \ref{tbl.ssim_sota_16} showed that the proposed Spach Transformer architecture observed the highest PSNR and SSIM performances, although the Spach Transformer had the least model parameters (19M) compared to the other reference methods (21M).
	Fig. \ref{fig.cnr_box} and \ref{fig.cnr_box_low} also showed that the Spach Transformer achieved the highest CNR performances.
	In Fig. \ref{fig.exp1}, \ref{fig.exp2}, \ref{fig.exp3}, and \ref{fig.exp4}, \textcolor{myCOLOR}{the adoption of the GCFN has shown that it helped to improve highlighting more of the tumor regions than excluding the GCFN.}
	We observed that the Spach Transformer could simultaneously do denoising and highlighting well.
	Besides that, although we trained the network architectures using $^{18}$F-FDG and $^{18}$F-ACBC datasets, the proposed Spach Transformer architecture was functioning well on $^{18}$F-DCFPyL and Ga-68 DOTATATE datasets as well as $^{18}$F-FDG and $^{18}$F-ACBC datasets \textcolor{myCOLOR}{due to similar local intensity patterns among different tracers.}
	
	\textcolor{myCOLOR}{We observed several limitations: In Table \ref{tbl.param_abl}, PSNR was increased, whereas SSIM was decreased when the proposed model's size grew. It may be that the best model was chosen based on PSNR for denoising purposes. Loss functions related to SSIM could potentially offer a viable approach to tackle this concern. While our primary focus has been on image quality metrics like PSNR and SSIM, one potential research direction to consider is the design of loss functions that incorporate other local features, such as radiomics of the lesions. \textcolor{myCOLOR2}{Delving into the equivalency of radiomics features is essential to understand rationales for possible false positive lesions, despite not identifying any in our analysis. We believe the lack of false positive lesions can be attributed to the high quality (e.g., alignment) of our input and reference data.} Moreover, building upon the promising PSNR and SSIM performances of the proposed method, our future research endeavors will be enhanced by incorporating the detection of lesions with low activity \textcolor{myCOLOR2}{to support the importance of evaluating the impact of intra-tumor heterogeneity \cite{hatt2009fuzzy, liu2021observer}}}
	
	\section{Conclusion}
	In this work, we proposed a spatial and channel-wise encoder-decoder transformer that utilized the spatial and channel information for PET image denoising. The proposed Spach Transformer achieved better performances in terms of CNR, PSNR, and SSIM on $^{18}$F-FDG,  $^{18}$F-ACBC,  $^{18}$F-DCFPyL, and $^{68}$Ga-DOTATATE datasets compared to other state-of-the-art methods. Our future work will focus on further evaluations using clinical datasets.
	
	\bibliography{references.bib}
	\bibliographystyle{IEEEtran}

	\newpage
	\onecolumn
	\appendix
	
	Additional details that are not included in the main text can be found in our appendices.
	In Appendix \ref{sec.A}, we provide additional information regarding our experiments on the 1/16 low-dose datasets, in addition to the experiments conducted on the 1/4 low-dose datasets.

	\subsection{Experiments on the 1/16 low-dose datasets} \label{sec.A}
\begin{figure*}[b!]
		\centering
		\begin{subfigure}[b]{\textwidth}
			\centering
			\includegraphics[width=0.9\textwidth]{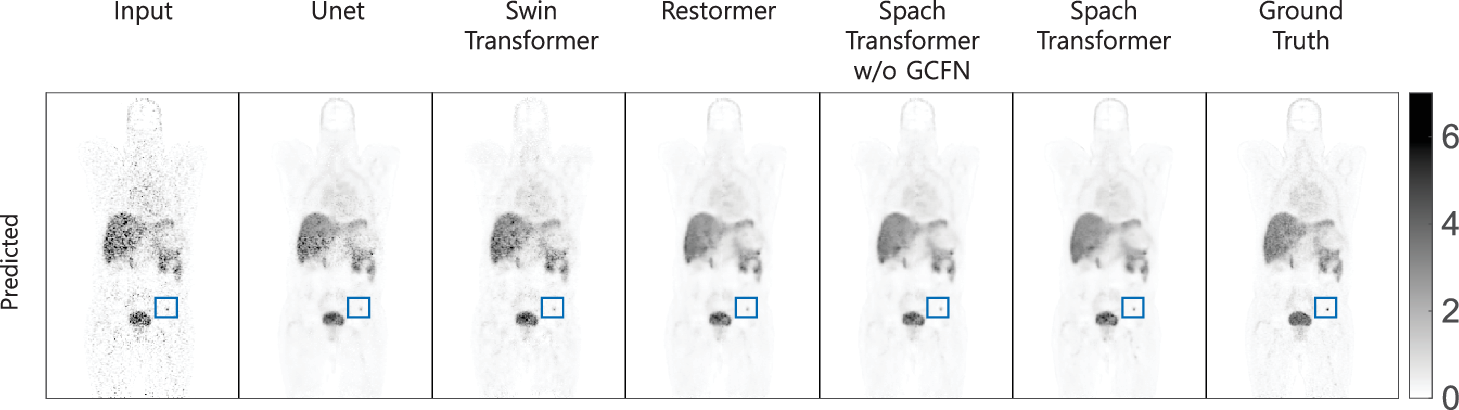}
			\caption{The predicted images and the distance images of the whole body.}
			\label{fig.exp1_1_low}
		\end{subfigure}
		\hfill
		\begin{subfigure}[b]{\textwidth}
			\centering
			\includegraphics[width=0.92\textwidth]{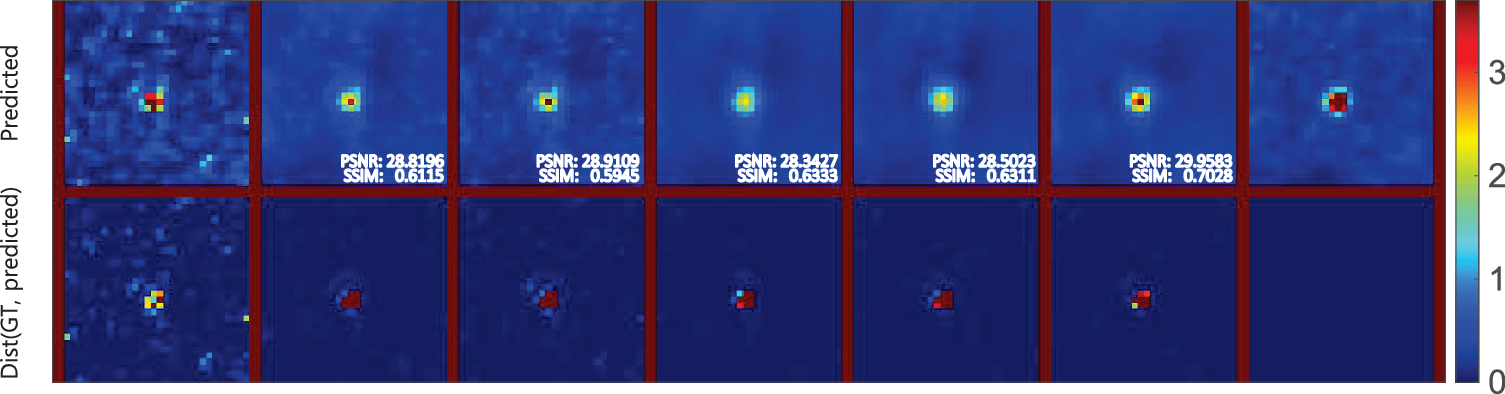}
			\caption{The enlarged images where the blue rectangulars are located in (a).}
			\label{fig.exp1_2_low}
		\end{subfigure}
		\caption{The denoised PET results of the state-of-the-art methods on a 1/16 dose $^{18}$F-DCFPyL dataset.}
		\label{fig.exp1_low}
	\end{figure*}

\begin{figure*}[b!]
		\centering
		\begin{subfigure}[b]{\textwidth}
			\centering
			\includegraphics[width=0.9\textwidth]{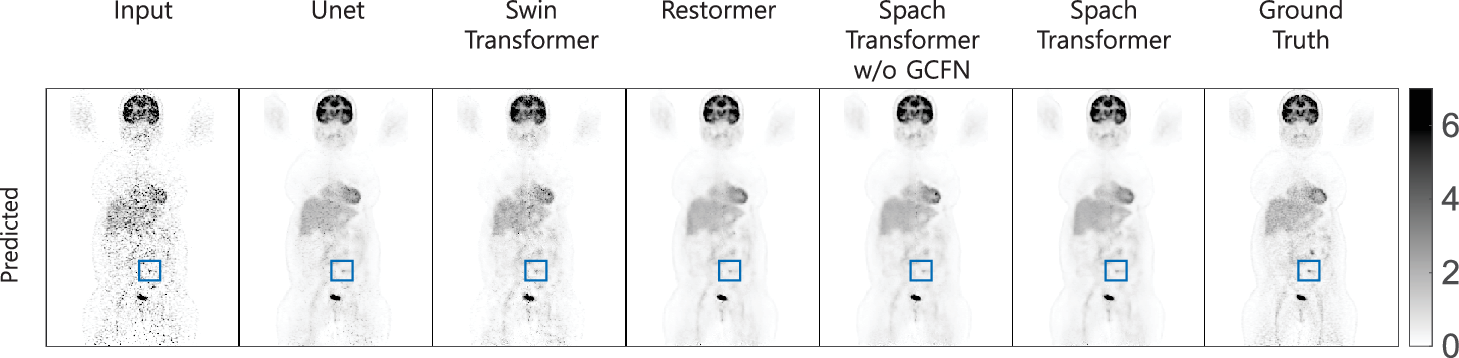}
			\caption{The predicted images and the distance images of the whole body.}
			\label{fig.exp2_1_low}
		\end{subfigure}
		\hfill
		\begin{subfigure}[b]{\textwidth}
			\centering
			\includegraphics[width=0.92\textwidth]{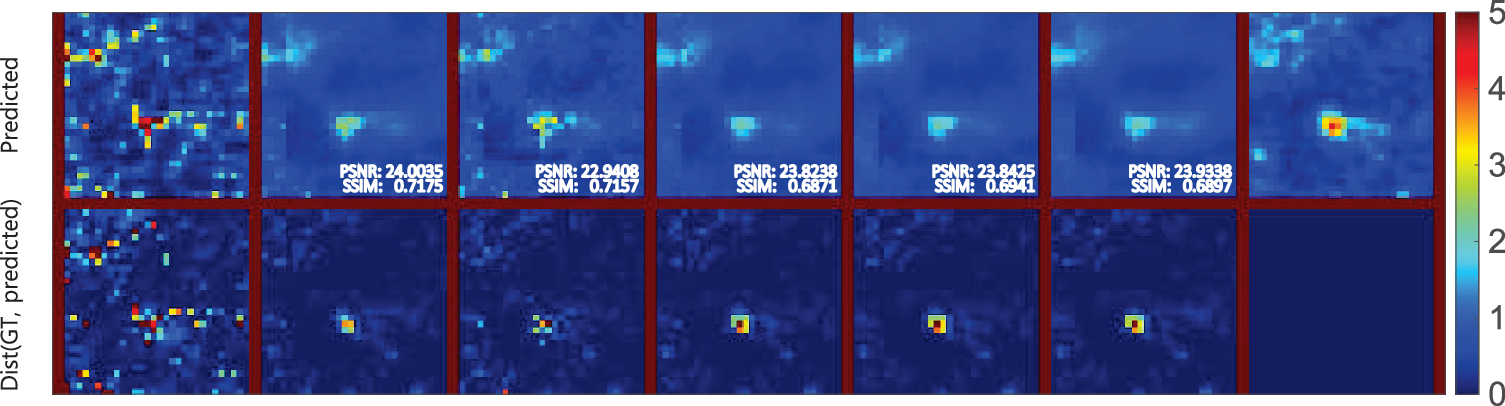}
			\caption{The enlarged images where the blue rectangulars are located in (a).}
			\label{fig.exp2_2_low}
		\end{subfigure} 
		\caption{The denoised PET results of the state-of-the-art methods on a 1/16 dose $^{18}$F-FDG dataset.}
		\label{fig.exp2_low}
	\end{figure*}

\begin{figure*}[t]
		\centering
		\begin{subfigure}[b]{\textwidth}
			\centering
			\includegraphics[width=0.9\textwidth]{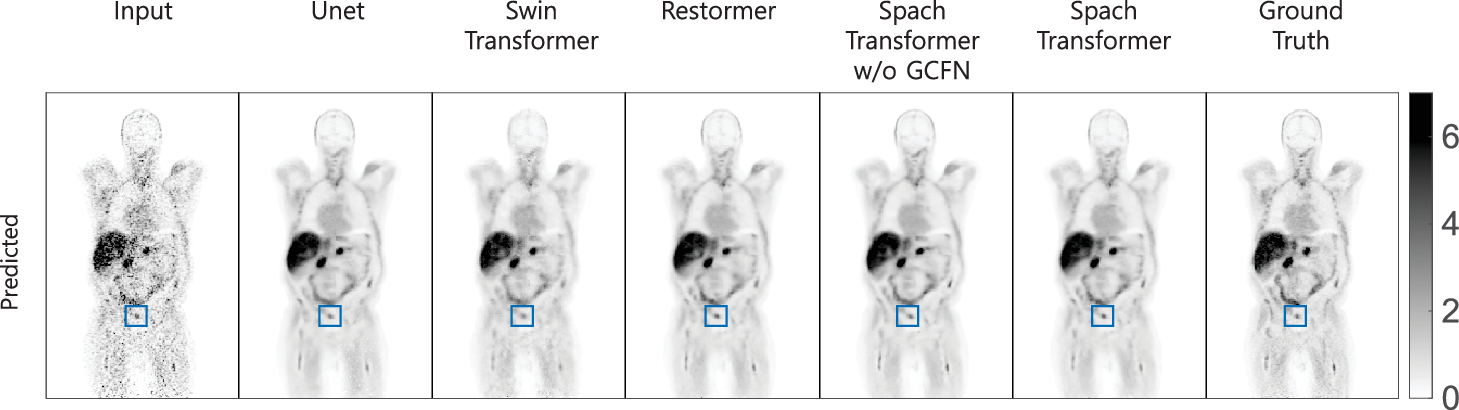}
			\caption{The predicted images and the distance images of the whole body.}
			\label{fig.exp3_1_low}
		\end{subfigure}
		\hfill
		\begin{subfigure}[b]{\textwidth}
			\centering
			\includegraphics[width=0.92\textwidth]{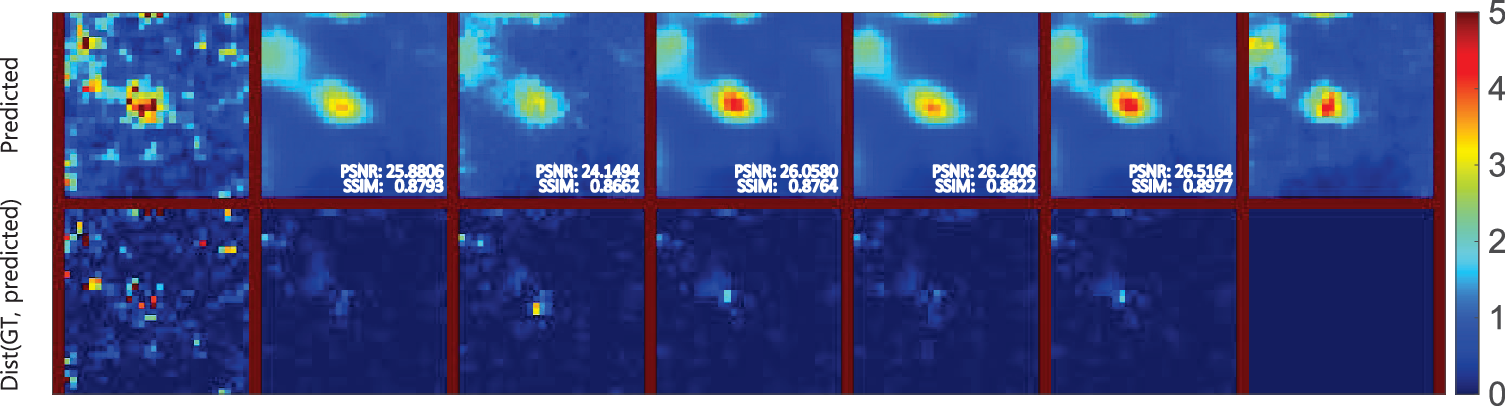}
			\caption{The enlarged images where the blue rectangulars are located in (a).}
			\label{fig.exp3_2_low}
		\end{subfigure}
		\caption{The denoised PET results of the state-of-the-art methods on a 1/16 dose $^{18}$F-ACBC dataset.}
		\label{fig.exp3_low}
	\end{figure*}

\begin{figure*}[t]
		\centering
		\begin{subfigure}[b]{\textwidth}
			\centering
			\includegraphics[width=0.9\textwidth]{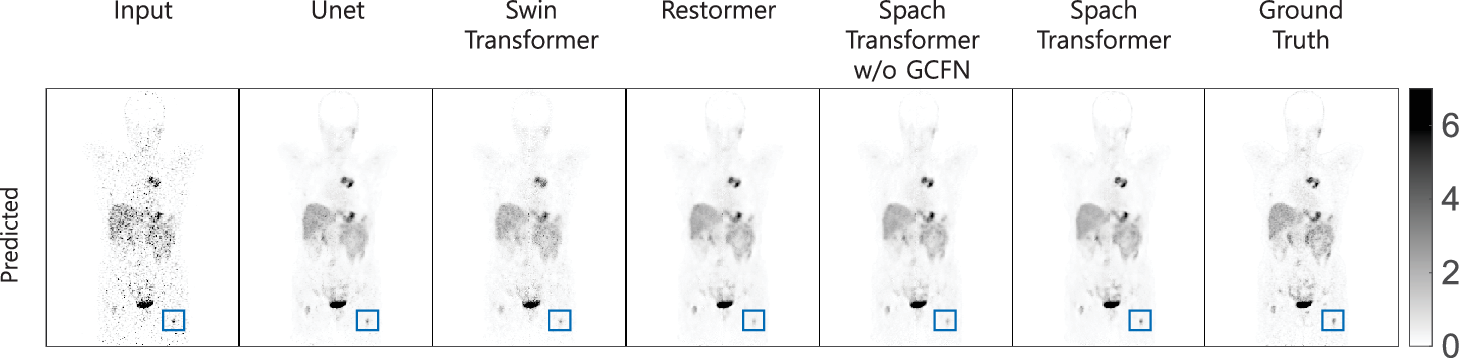}
			\caption{The predicted images and the distance images of the whole body.}
			\label{fig.exp4_1_low}
		\end{subfigure}
		\hfill
		\begin{subfigure}[b]{\textwidth}
			\centering
			\includegraphics[width=0.92\textwidth]{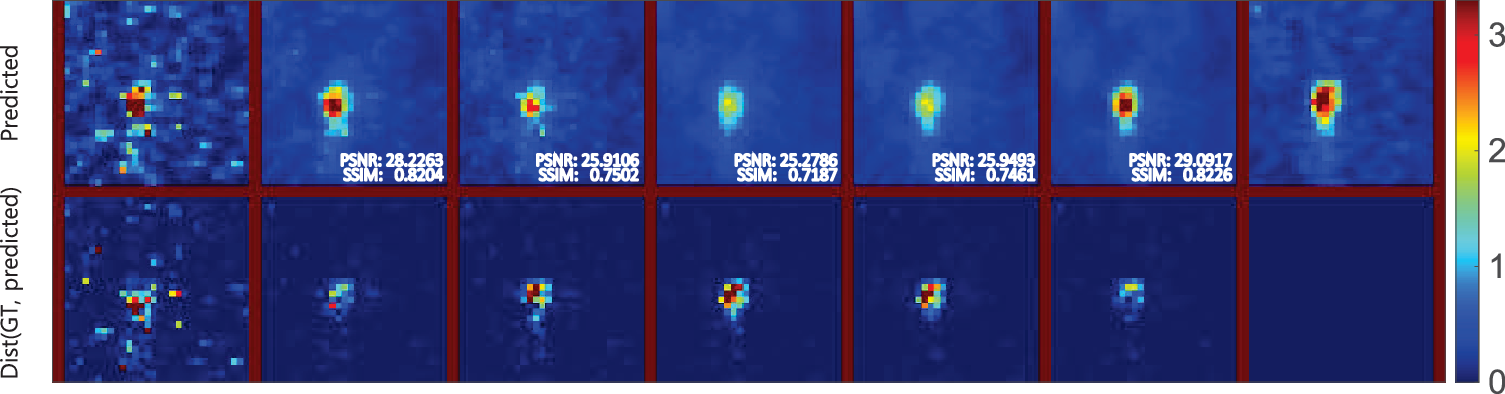}
			\caption{The enlarged images where the blue rectangulars are located in (a).}
			\label{fig.exp4_2_low}
		\end{subfigure}
		\caption{The denoised PET results of the state-of-the-art methods on a 1/16 dose Ga-68 DOTATATE dataset.}
		\label{fig.exp4_low}
	\end{figure*}

	Fig. \ref{fig.exp1_low}, \ref{fig.exp2_low}, \ref{fig.exp3_low}, and \ref{fig.exp4_low} illustrate the results obtained from the Spach Transformer and the reference methods using the 1/16 dose $^{18}$F-DCFPyL, $^{18}$F-FDG, $^{18}$F-ACBC, and Ga-68 DOTATATE datasets.
	In Fig. \ref{fig.exp1_1_low}, \ref{fig.exp2_1_low}, \ref{fig.exp3_1_low}, and \ref{fig.exp4_1_low}, we observe that the Restormer outperforms the Unet and Swin Transformer in terms of denoising performance.
	On the other hand, Fig. \ref{fig.exp1_2_low}, \ref{fig.exp2_2_low}, \ref{fig.exp3_2_low}, and \ref{fig.exp4_2_low} demonstrate that the Unet and Swin Transformer exhibit relatively higher uptake values than the Restormer. In the $^{18}$F-FDG dataset, the Swin Transformer exhibited a superior uptake compared to all the compared methods. However, it also generated the most noised prediction among the methods.

\end{document}